\documentclass[12pt]{article}
\usepackage{latexsym}
\usepackage{amssymb, amsmath, xspace, lscape,  latexsym}
\usepackage{amsthm}
\usepackage{graphicx}
\usepackage{tikz}

\newcommand{\bea}{\begin{eqnarray}}
\newcommand{\eea}{\end{eqnarray}}
\def\beq#1#2\eeq{
        \begin{equation}
        \label{#1}
            #2
        \end{equation}}

\newcommand{\al}{\alpha}
\newcommand{\la}{\lambda}

\newcommand{\C}{\mathbb C}

\newcommand{\up}{\upsilon}
\newcommand{\va}{\varepsilon}
\newcommand{\rw}{\rightarrow}
\newcommand{\ep}{\epsilon}

\renewcommand{\hat}{\widehat}
\renewcommand{\tilde}{\widetilde}

\def\btheor#1\etheor{
        \begin{theor}
            #1
        \end{theor}
    }

    \def\bsled#1\esled{
        \begin{sled}
            #1
        \end{sled}   }

\newcommand{\bq}{\begin{equation}}
\newcommand{\eq}{\end{equation}}
\newcommand{\bbq}{\begin{equation*}}
\newcommand{\eeq}{\end{equation*}}

\newtheorem{rem}{Remark}

\newtheorem{theorem}{Theorem}

\newtheorem{lemma}{Lemma}
\newtheorem{prop}{Proposition}

\def\hm#1{#1\nobreak\discretionary{}{\hbox{\m@th$#1$}}{}}
\def\mi#1{\discretionary{\hbox{\m@th$#1$}}{\hbox{\m@th$#1$}}{}}

\vspace{4ex}
\begin{document}
\baselineskip=20pt
\title{\bf Critical edge behavior in the perturbed Laguerre ensemble and the Painlev\'e \uppercase\expandafter{\romannumeral5} transcendent}
\author{Min Chen$^{1}$, Yang Chen$^{2}$, En-Gui Fan$^{1}$\thanks{Corresponding author: faneg@fudan.edu.cn}\\
{\normalsize $^{1}$School of Mathematical Science, Fudan University, Shanghai 200433, P.R. China}\\
{\normalsize $^{2}$Faculty of Science and Technology, Department of Mathematics, University of Macau}
       }
\date{\today}
\maketitle
\begin{abstract}
\baselineskip=20pt

In this paper, we consider  the perturbed Laguerre unitary ensemble described by the weight function of
$$w(x,t)=(x+t)^{\la}x^{\al}e^{-x}$$
with $ x\geq 0,\ t>0,\  \al>0,\ \al+\la+1>0.$
The  Deift-Zhou nonlinear steepest descent approach  is used to analyze  the limit of the eigenvalue correlation kernel.
It was found that under the double scaling $s=4nt,$ $n\rw \infty,$ $t\rw 0 $ such that $s$ is positive and finite, at the hard edge, the limiting kernel can be described by the $\varphi$-function related to a
third-order nonlinear differential equation, which is equivalent to a particular Painlev\'e ${\rm \uppercase\expandafter{\romannumeral5}}$ (shorted as ${\rm P_{\uppercase\expandafter{\romannumeral5}}}$) transcendent via a simple transformation. Moreover, this ${\rm P_{\uppercase\expandafter{\romannumeral5}}}$ transcendent is equivalent to a  general Painlev\'e ${\rm \uppercase\expandafter{\romannumeral3}}$ transcendent. For large $s,$ the ${\rm P_{\uppercase\expandafter{\romannumeral5}}}$ kernel reduces to the Bessel kernel $\mathbf{J}_{\al+\la}.$ For small $s,$ the ${\rm P_{\uppercase\expandafter{\romannumeral5}}}$ kernel reduces to another Bessel kernel $\mathbf{J}_{\al}.$
At the soft edge, the limiting kernel is the Airy kernel as the classical Laguerre weight.

\end{abstract}
\noindent

\newpage

\setcounter{equation}{0}
\section{Introduction and statement of results}

\par
In random matrix theory, the unitary random matrix ensemble on the space of $n\times n$ positive definite Hermitian matrices, described by the following measure,
\bbq\label{R2}
Z_{n}^{-1}e^{{\rm Tr}\log w\left(M\right)}dM, \;\;{\rm and}\;\;dM=\prod_{j=1}^{n}dM_{jj}\prod_{1\leq\ell<k\leq n}d{\rm Re}M_{\ell{k}}d{\rm Im}M_{\ell{k}},
\eeq
where the normalization constant $Z_{n}$ ensures the above measure is a probability measure, and $w(x)$ is the weight function on $\mathcal{L}\subset \mathbb{R}.$
\par
The correlation kernel in the following as an important object is investigated in random matrix theory \cite{DeiftBook, D1970, F2010, M2004},
\bq\label{00a1}
K_{n}(x,y)=\left(w(x)\right)^{\frac{1}{2}}\left(w(y)\right)^{\frac{1}{2}}\sum_{j=0}^{n-1}\frac{\pi_{j}(x)\pi_{j}(y)}{h_{j}},
\eq
where
\bq\label{T6}
\int_{\mathcal{L}}\pi_{n}(x)\pi_{m}(x)w(x)dx=h_{n}\delta_{nm},
\eq
and $\pi_{n}(x)$  is the monic orthogonal polynomials associated with the weight $w(x)$ on $\mathcal{L}.$
\par
It is interesting to study the local eigenvalue behavior by characterizing the kernel in the large $n$ limit with a suitable scale. For instance, in the case of Laguerre Unitary Ensemble (LUE), see \cite{ B1965, F2010, Nw1991}, the limiting density of eigenvalue for a fixed $x,$ is known as a type of Mar\v{c}enko-Pastur law \cite{MP1967} as follows
\bq\label{R24}
\mu(x)=\lim_{n\rw \infty}4K_{n}\left(4nx,4nx\right)=\frac{2}{\pi}\sqrt{\frac{1-x}{x}},\;\; 0<x<1.
\eq
In previous works \cite{F2003,TW1994,TW1993AK}, at the hard edge of the eigenvalue density, the limiting kernel is known as the Bessel Kernel
\bbq\label{R03}
\mathbf{J}_{\beta}(x,y):=\frac{J_{\beta}(\sqrt{x})\sqrt{y}J'_{\beta}(\sqrt{y})-\sqrt{x}J'_{\beta}(\sqrt{x})J_{\beta}(\sqrt{y})}{2(x-y)},
\eeq
and at the soft edge of the eigenvalue density, the limiting kernel is known as the Airy Kernel
\bbq\label{R02}
\mathbf{A}(x,y):=\frac{Ai(x)Ai'(y)-Ai'(x)Ai(y)}{x-y}.
\eeq
\par
Since the Deift-Zhou nonlinear steepest descent approach applied in random matrix theory, some new properties has been identified in random matrix theory. For example, the limiting kernels in the bulk of the spectrum is usually described by the sine kernel, which is well-known as universality phenomenon (see \cite{DeiftBook, DX1999, DKV1999, KV2002, V2007}). Using this powerful method, it has been found that some limiting kernels are related to Painlev\'e equations. In a particular double scaling scheme, the limiting kernel involves Painlev\'e \uppercase\expandafter{\romannumeral1} equation in \cite{CV2007}. An appropriate double scaling limit of the
kernel relates to the Painlev\'e \uppercase\expandafter{\romannumeral2} equation, see \cite{BI1999, CK2008, C2008K}. A limiting kernel relates to the Painlev\'e \uppercase\expandafter{\romannumeral2} equation and $4\times4$ RH problem discussed in the Hermitian two matrix model, see \cite{D2014}. An $\al$-generalized Airy kernel expressed by a solution of a Painlev\'e \uppercase\expandafter{\romannumeral34} equation in \cite{IKO2008}. A Painlev\'e \uppercase\expandafter{\romannumeral3} equation involves the double scaling limit of the kernel, and this ${\rm P_{\uppercase\expandafter{\romannumeral3}}}$ kernel translates to Bessel kernel and Airy kernel in certain conditions \cite{XDZ2014}.
\par
Moreover, a general Jacobi unitary ensemble described by the weight function $w(x)=(1-x)^{\al}(1+x)^{\beta}h(x),$ $\al>-1,$ $\beta>-1,$ where $h(x)$ is positive and analytic on $[-1, 1].$ The university at the hard edge is studied in \cite{KV2002, KM2004}. Under certain double scaling scheme, the limiting kernel can be expressed in terms of a solution of Painlev\'e \uppercase\expandafter{\romannumeral5} function and this ${\rm P_{\uppercase\expandafter{\romannumeral5}}}$ kernel degenerates to Bessel kernels with different conditions, see \cite{XZ2015}. A limiting kernel in terms of the hypergeometric functions investigated in \cite{DKV2010}. A type of general Bessel kernel derives from a scaling limit of kernel at the hard edge and its explicit integration formula is given in \cite{KZ2014}.
\par
In this paper, we consider the following weight function
\bq\label{R1}
w(x):=w(x,t)=(x+t)^{\la}x^{\al}e^{-x},\;\; x\geq0,\;\; t>0,\;\; \al>0,\;\; \al+\la+1>0.
\eq
When $t=0,$ or $\la=0,$ the weight degenerates to the classical Laguerre weight. A more general weight of this type arises from Multiple-Input-Multiple-Output (MIMO) wireless communication system. It is interesting that the free parameter $\la$ ``generates" the Shannon capacity, and the moment generating function (MGF) can be expressed by the ration of Hankel determinants \cite{YangMcky2012} for fine $n,$ where the Hankel determinant involves a particular Painlev\'e ${\rm \uppercase\expandafter{\romannumeral5}}$ equation. The double scaling limit of the Hankel determinant described by another particular ${\rm P_{\uppercase\expandafter{\romannumeral5}}}$ equation which is equivalent to a particular ${\rm P_{\uppercase\expandafter{\romannumeral3}}},$ see \cite{CBC2016}. For more information related to wireless communication, we refer \cite{BasorChen, YangHaqMcky2013} and references therein. Moreover, a special case of (\ref{R1}) as $w(x, t)=(x+t)^{a}x^{2}e^{-x},$ $x>0,$ $a>-1$ appeared in the study of the smallest eigenvalue at the hard edge of the Laguerre unitary ensemble, see \cite{FW2007}.
\par
The weight (\ref{R1}) can be seen as $x^{\la+\al}e^{\frac{\lambda{t}}{x}-x}.$ Up to an essential singular point, this is the singularly perturbed Laguerre weight considered in \cite{ChenIts2010}, and they obtained a connection with ${\rm P_{\uppercase\expandafter{\romannumeral3}}}$ for finite $n$.
Heuristically, this maybe seen as follows (note $s=4nt$),
\bbq
(t+x)^{\la}x^{\alpha}e^{-x}
=x^{\alpha+\lambda}\left(\left(1+s/4nx\right)^{\frac{4nx}{s}}\right)^{\frac{\la{s}}{4nx}}\; e^{-x}\to x^{\lambda+\alpha}
e^{\frac{\la{t}}{x}-x},\;\;n\to\infty.
\eeq
By the Riemann-Hilbert approach, the double scaling limit of the kernel associated with this singular weight presents as a ${\rm P_{\uppercase\expandafter{\romannumeral3}}}$ kernel, see \cite{XDZ2014}, with physical background provided in \cite{OVAKE2007} and for a type of singularly perturbed Gaussion weight relevant study (for Hermite) see \cite{BMM2014}.
\par
It is interesting to investigate the double scaling limit of the kernel related with the weight (\ref{R1}) on $(0,\infty).$ We adapt Deift-Zhou nonlinear steepest descent method to investigate the limiting behavior of the kernel in this paper.
\par
For preparation, we recall that a Lax pair in the following is given by Kapaev and Hubert \cite{KH1999} and note that this Lax pair and its Painlev\'e ${\rm \uppercase\expandafter{\romannumeral5}}$ equation do not contain in \cite{FIK2006}, with a change of variables $\lambda$ and $x$ in \cite{KH1999} as $\xi$ and $s,$ respectively, which will help us to derive ${\rm P_{\uppercase\expandafter{\romannumeral5}}}$ transcendent directly in our situation.
\begin{prop}(Kapaev and Hubert \cite{KH1999})
The Lax pair for $\Psi$ is given by
\bq\label{Rr1}
\Psi_{\xi}\Psi^{-1}=A=\left(
\begin{matrix}
0&1\\
0&0\\
\end{matrix}
\right)
+\frac{1}{\xi}\left(
\begin{matrix}
a&b\\
c&-a\\
\end{matrix}
\right)
+\frac{1}{\xi-s}\left(
\begin{matrix}
p&q\\
r&-p\\
\end{matrix}
\right),
\eq
\bq\label{Rr2}
\Psi_{s}\Psi^{-1}=U=-\frac{1}{\xi-s}\left(
\begin{matrix}
p&q\\
r&-p\\
\end{matrix}
\right).
\eq
where $a,$ $b,$ $c,$ $p,$ $q,$ and $r$ are dependent on $s$ and $\rho,$ $\mu$ and $\upsilon$ are constant. The compatibility condition for the system (\ref{Rr1}) and (\ref{Rr2}) is equivalent to the system of equations
\bbq
b=(a^{2}-\mu^{2})\frac{y-1}{\rho},\;\;c=-\frac{\rho}{y-1},\;\;p=\frac{s}{2}\frac{y_{s}}{y-1}-ay,\;\;q=-\left(p^{2}-\upsilon^{2}\right)\frac{y-1}{\rho{y}},\;\; r=\frac{\rho{y}}{y-1},
\eeq
where the unknown function $a(s)$ satisfies the differential equation,
\bbq\label{REE3}
a_{s}+a\frac{y_{s}}{y-1}=\frac{s}{4}\frac{(y_{s})^{2}}{y(y-1)^{2}}+\frac{1}{s}\left(\mu^{2}y-\frac{\upsilon^{2}}{y}\right),
\eeq
and the function $y(s)$ satisfies the special ${\rm P_{\uppercase\expandafter{\romannumeral5}}}(2\mu^{2}, -2\upsilon^{2}, 2\rho, 0)$ as follows
\bq\label{REE4}
y_{ss}=\left(\frac{1}{2y}+\frac{1}{y-1}\right)y_{s}^{2}-\frac{y_{s}}{s}+\frac{2(y-1)^{2}}{s^{2}}\left(\mu^{2}y-\frac{\upsilon^{2}}{y}\right)+\frac{2\rho}{s}y.
\eq
Moreover, the above ${\rm P_{\uppercase\expandafter{\romannumeral5}}}$ equation is equivalent to the general ${\rm P_{\uppercase\expandafter{\romannumeral3}}}$ equation \cite{GLS2002}.
\end{prop}
\par
In our case, the following Lax pair for $\Phi(\xi, s)$ and an auxiliary function $r(s)$ is a solution of a third-order nonlinear differential equation associated with a particular Painlev\'e ${\rm \uppercase\expandafter{\romannumeral5}}$ transcendent.

\subsection{Lax pair, a third-order nonlinear differential equation and ${\rm P_{\uppercase\expandafter{\romannumeral5}}}$}

\begin{prop}
The Lax pair for $\Phi(\xi, s)$ is given by
\bq\label{RE24}
\Phi_{\xi}(\xi,s)=\left(A_{0}(s)+\frac{A_{1}(s)}{\xi}+\frac{A_{2}(s)}{\xi-s}\right)\Phi(\xi,s),
\eq
\bq\label{RE25}
\Phi_{s}(\xi,s)=\frac{B_{2}(s)}{\xi-s}\Phi(\xi,s),
\eq
where
\bq\label{RE26}
A_{0}(s)=\frac{1}{2}\left(
\begin{matrix}
0&0\\
i&0\\
\end{matrix}
\right),\;
A_{1}(s)=\left(
\begin{matrix}
-\frac{1}{4}+\frac{1}{2}r(s)+q'(s)&-i\left(\frac{1}{2}+r'(s)\right)\\
i\left(-q(s)+t'(s)\right)&\frac{1}{4}-\frac{1}{2}r(s)-q'(s)\\
\end{matrix}
\right),\;
A_{2}(s)=-B_{2}(s),
\eq
\bq\label{RE27}
B_{2}(s)=\left(
\begin{matrix}
q'(s)&-ir'(s)\\
it'(s)&-q'(s)\\
\end{matrix}
\right),
\eq
and $q(s),$ $q'(s)$ and $t'(s)$ in terms of $r(s)$ and its derivatives,
\bq\label{RE28}
q'=-sr''+r'r-\frac{1}{2}r',
\eq
\bq\label{RE29}
t'=\frac{\lambda^{2}-(2sr''-2r'r+r')^2}{4r'},
\eq
\bq\label{RE30}
q=-\frac{8s^2r''^2-r'(4r(r-1)(2r'+1)+2r'+4\lambda^2-4\alpha^2+1)-2\lambda^2}{8r'(2r'+1)},
\eq
with $\al>0,$ $\al+\la+1>0.$

\end{prop}

\begin{rem}
It's easy to check that the Lax pair (\ref{Rr1}), (\ref{Rr2}) for $\Psi$ matches the Lax pair (\ref{RE24}), (\ref{RE25}) for $\Phi$, in the sense of the following linear transformation
\bbq\label{RE32}
\Phi=\left(
\begin{matrix}
0&1\\
\frac{i}{2}&0\\
\end{matrix}
\right)\Psi.
\eeq
From the above equation, we can identify that $\mu^{2}=\frac{\alpha^{2}}{4},$ $\upsilon^{2}=\frac{\lambda^{2}}{4},$ and $\rho=\frac{1}{4},$ and especially the unknown function $a(s)=\frac{1}{4}+\frac{1}{2}r(s)-q'(s).$ Then, we can identify all other quantities in the Lax pair can be expressed in terms of the auxiliary function $r(s)$ and its derivatives.
\par
Although the ${\rm P_{\uppercase\expandafter{\romannumeral5}}}$ transcendent in (\ref{REE4}) is equivalent to the general ${\rm P_{\uppercase\expandafter{\romannumeral3}}}$ transcendent \cite{GLS2002}, but there is no algebraic gauge transformation from the Lax pair for ${\rm P_{\uppercase\expandafter{\romannumeral5}}}$ equation to the Lax pair for ${\rm P_{\uppercase\expandafter{\romannumeral3}}}$ equation which has only two irregular singular points of zero and infinity.
\end{rem}

\begin{prop}
If $r(s)$  satisfies the Lax pair (\ref{RE24}) and (\ref{RE25}) for $\Phi(\xi,s),$ then it also satisfies a third-order nonlinear differential equation,
\bq\label{RE31}
8s^2r'(2r'+1)r'''-4s^2(1+4r')r''^2+8sr'(2r'+1)r''-4sr'^2(2r'+1)^2+\lambda^2(2r'+1)^2-4\alpha^2r'^{2}=0,
\eq
where $\al>0,$ $\al+\la+1>0,$ and $r(s)$ satisfies the following boundary conditions,
\bq\label{R7}
r(0)=\frac{1-4(\alpha+\lambda)^{2}}{8},\;\;{\rm and}\;\; r(s)=-\la{s^{-\frac{1}{2}}}+\mathcal{O}\left(s^{-1}\right),\;\;s\rw \infty.
\eq
If $\al+\la>0$ and $s\rw 0,$ then
\bq\label{T26}
r'(0)=-\frac{\la}{2(\al+\la)}.
\eq
\end{prop}

\begin{proof}
With the aid of the equations (\ref{RE28}) and (\ref{RE30}) which derive form the compatibility condition of the Lax pair (\ref{RE24}) and (\ref{RE25}) for $\Phi(\xi,s),$ one takes derivative of (\ref{RE30}) and a combination of (\ref{RE28}), then one obtains the third-order nonlinear differential equation (\ref{RE31}). The boundary conditions of $r(s)$ in (\ref{R7}) derives from (\ref{RE74}), (\ref{R10}) for $s\rw 0$ and $s\rw \infty,$ respectively. Moreover, (\ref{R3}) gives (\ref{T26}).
\end{proof}

\begin{rem}
The third-order equation (\ref{RE31}) is an integrable equation. The integration of the equation (\ref{RE28}) presents as
\bq\label{RRE03}
q(s)=-sr'(s)+\frac{r(s)^{2}+r(s)}{2}+c_{1},
\eq
where $c_{1}$ is an integration constant. From the boundary condition (\ref{RE74}), it follows that $c_{1}=0$ in this situation. By the equivalence of $q(s)$ in (\ref{RE30}) and (\ref{RRE03}), then one finds $r(s)$ also satisfies another second-order nonlinear differential equation,
\bq\label{RRE04}
s^2r''^{2}-2sr'^{3}+\left(2r+2c_{1}-s-\frac{1}{4}\right)r'^{2}+\left(r+\frac{\alpha^{2}}{2}+c_{1}-\frac{1}{8}-\frac{\lambda^{2}}{2}\right)r'-\frac{\lambda^{2}}{4}=0,
\eq
it can also be rewritten as follows
\bq\label{RRE05}
\frac{s^{2}r''^{2}}{r'(2r'+1)}-\frac{2sr'^{2}}{2r'+1}-\frac{sr'}{2r'+1}+\frac{\alpha^{2}-\lambda^{2}}{2(2r'+1)}-\frac{\lambda^{2}}{4r'(2r'+1)}+r=\frac{1}{8}-c_{1},
\eq
and one takes derivative with respective to $s$ on both sides of the above equation, then one obtains the third-order equation (\ref{RE31}). Taking the inverse procedure, the third-order equation in (\ref{RE31}) is indeed an integrable equation.
\par
Furthermore, inserting the equation (\ref{RRE03}) into (\ref{RE43}), then one finds $r(s)$ satisfies the following equation
\bq\label{RRE06}
8s^2r'r'''-4s^2r''^{2}+8sr'r''-16sr'^{3}+\left(8r-4s-1+8c_{1}\right)r'^{2}+\lambda^{2}=0.
\eq
Then the third-order equation (\ref{RE31}) is the sum of $-8r'(s)$ times of the equation (\ref{RRE04}) and $2r'(s)+1$ multiples the equation (\ref{RRE06}).
\end{rem}

\begin{prop}
Let
\bq\label{R8}
r'(s)=\frac{y(s)}{2(1-y(s))},
\eq
then $y(s)$ satisfies the following ${\rm P_{\uppercase\expandafter{\romannumeral5}}}\left(\frac{\alpha^{2}}{2},\frac{\lambda^{2}}{2},\frac{1}{2},0\right)$ transcendent,
\bq\label{RE33}
y''(s)=\left(\frac{1}{2y(s)}+\frac{1}{y(s)-1}\right)y'^{2}(s)-\frac{y'(s)}{s}+\frac{(y(s)-1)^{2}}{2s^{2}}\left(\alpha^{2}y(s)-\frac{\lambda^{2}}{y(s)}\right)+\frac{y(s)}{2s},
\eq
where $\al>0,$ $\al+\la+1>0,$ and $y(s)$ satisfies the boundary conditions
\bq\label{R98}
y(0)=-\frac{\la}{\al},\;\; {\rm and}\;\; y(s)=-\la{s^{-\frac{1}{2}}}+\mathcal{O}\left(s^{-1}\right),\;\; s\rw \infty.
\eq
\end{prop}

\begin{proof}
Inserting (\ref{R8}) into the third-order nonlinear differential equation (\ref{RE31}), one finds the ${\rm P_{\uppercase\expandafter{\romannumeral5}}}$ transcendent in (\ref{RE33}). A combination of $(\ref{R8})$ and $(\ref{R7})$ yields the initial data (\ref{R98}) for $y(s).$
The above ${\rm P_{\uppercase\expandafter{\romannumeral5}}}$ equation is also derived by the analysis of the single-user MIMO system \cite{CBC2016}, but it is not from the Riemann-Hilbert approach.
\end{proof}

\subsection{Main results}
\par
We denote the kernel (\ref{00a1}) associated with the weight (\ref{R1}) as $K_{n}(x,y;t)$ in the rest of this paper. By two scaling steps, $K_{n}(x,y;t)$ scaled as $4nK_{n}(4nx,4ny;t),$ and the ``coordinates'' $x$ and $y$ are re-scaled as $x=\frac{u}{16n^{2}},$ $y=\frac{{\rm v}}{16n^{2}}.$ After that, the limiting kernel is described by the $\varphi-$functions which involve the above ${\rm P_{\uppercase\expandafter{\romannumeral5}}}$ equation (\ref{RE33}), also known as ${\rm P_{\uppercase\expandafter{\romannumeral5}}}$ kernel.
\begin{theorem}
Let $s=4nt,$  $n\rw \infty,$  and $t\rw 0^{+},$ such that $s$ is positive and finite, then
\bq\label{T1}
\lim_{n\rw \infty}\frac{1}{4n}K_{n}(\frac{u}{4n},\frac{{\rm v}}{4n}; \frac{s}{4n})=\frac{\varphi_{1}(-{\rm v},s)\varphi_{2}(-u,s)-\varphi_{1}(-u,s)\varphi_{2}(-{\rm v},s)}{i2\pi(u-{\rm v})},
\eq
which is uniform for $u,{\rm v}\in (0,\infty),$ $\varphi_{k}(\xi,s),k=1,2$ satisfy the following second order differential equation,
\bq\label{R11}
\varphi''(\xi)+\left(\frac{1}{\xi}+\frac{1}{\xi-s}-\frac{1}{\xi-s-2sr'}\right)\varphi'(\xi)+W(\xi,s)\varphi(\xi)=0,
\eq
where $\xi\in \widehat{\Omega}_{3},$ see Figure 1, $\varphi'(\xi)$ and $r'$ denote $d\varphi(\xi,s)/d{\xi},$ $d{r(s)}/d{s},$ respectively, $r'(s)$ involves ${\rm P_{\uppercase\expandafter{\romannumeral5}}}$ (see (\ref{R8}) and (\ref{RE33})) and $W(\xi,s)$ is given by
\begin{align*}\label{R12}
W(\xi,s)=&-\frac{(s+2sr'-\xi)[4\xi(\xi-s)r'^{2}+2((s-\xi)(\al^{2}+\xi)+\la^{2}\xi)r'+\la^{2}\xi]}{8\xi^{2}(\xi-s)^{2}r'(1+2r')(s+2sr'-\xi)}\nonumber\\
&+\frac{4s^{2}(s-\xi)\xi{r''}[r''(s-\xi+2sr')+2r'(1+2r')]}{8\xi^{2}(\xi-s)^{2}r'(1+2r')(s+2sr'-\xi)},
\end{align*}
and $\al>0,$ $\al+\la+1>0.$ If $s=0$ and $\al+\la>0,$ then the equation (\ref{R11}) degenerates to a modified Bessel differential equation
\bq\label{R13}
\varphi''(\xi)+\frac{1}{\xi}\varphi'(\xi)-\left(\frac{1}{4\xi}+\frac{(\al+\la)^{2}}{4\xi^{2}}\right)\varphi(\xi)=0.
\eq
\end{theorem}
\par
The ${\rm P_{\uppercase\expandafter{\romannumeral5}}}$ kernel (\ref{T1}) degenerates to the Bessel kernel $\mathbf{J}_{\al+\la}$ as $s\rw 0.$ 
\begin{theorem}
Let $s=4nt,$ $n\rw \infty,$ and $t\rw 0^{+},$ such that $s\rw 0^{+},$ then the following limiting kernel can be identified as the Bessel kernel $\mathbf{J}_{\al+\la}$,
\bq\label{T8}
\lim_{n\rw \infty}\frac{1}{4n}K_{n}(\frac{u}{4n},\frac{{\rm v}}{4n}; \frac{s}{4n})=\mathbf{J}_{\al+\la}(u,{\rm v})=\frac{J_{\al+\la}(u){\rm v}J'_{\al+\la}({\rm v})-J_{\al+\la}({\rm v})uJ'_{\al+\la}(u)}{2(u-{\rm v})},
\eq
where $\al>0,$ $\al+\la+1>0.$ The scaling limit of the kernel is independent of $s$ and uniform for $u, {\rm v}\in (0,\infty).$
\end{theorem}
The ${\rm P_{\uppercase\expandafter{\romannumeral5}}}$ kernel (\ref{T1}) degenerates to the Bessel kernel $\mathbf{J}_{\al}$ as $s\rw \infty.$
\begin{theorem}
Let $s=4nt,$ $t\in (0,c],$ $c>0,$ $n\rw \infty,$ and $t\rw c^{-},$ such that $s\rw \infty,$ then the following limiting kernel is explicit as the Bessel kernel $\mathbf{J}_{\al}$,
\bq\label{T9}
\lim_{n\rw \infty}\frac{1}{4n}K_{n}(\frac{u}{4n},\frac{{\rm v}}{4n}; \frac{s}{4n})=\mathbf{J}_{\al}(u,{\rm v})=\frac{J_{\al}(u){\rm v}J'_{\al}({\rm v})-J_{\al}({\rm v})uJ'_{\al}(u)}{2(u-{\rm v})},
\eq
where $\al>0,$ $\al+\la+1>0.$ The scaling limit of the kernel is independent of $s$ and uniform for $u, {\rm v}\in (0,\infty).$
\end{theorem}

The limiting behavior of the kernel at the soft edge. Firstly, $K_{n}(x,y;t)$ scaled as $4nK_{n}(4nx,4ny;t).$ Secondly, $x$ and $y$ scaled as $x=1+(2n)^{-\frac{2}{3}}u,$ $y=1+(2n)^{-\frac{2}{3}}{\rm v}.$ After that the limiting kernel reads as the following Airy kernel.
\begin{theorem}
The kernel (\ref{00a1}) associated with the weight (\ref{R1}) denotes as $K_{n}(x,y;t)$. If $x$ and $y$ scaled as $x=4n+2(2n)^{\frac{1}{3}}u,$ $y=4n+2(2n)^{\frac{1}{3}}{\rm v},$ then the limiting kernel is the Airy kernel $\mathbf{A},$
\bq\label{T10}
\lim_{n\rw \infty}2(2n)^{\frac{1}{3}}K_{n}(4n+2(2n)^{\frac{1}{3}}u,4n+2(2n)^{\frac{1}{3}}{\rm v}; t)=\mathbf{A}(u,{\rm v})=\frac{Ai(u)Ai'({\rm v})-Ai({\rm v})Ai'(u)}{u-{\rm v}},
\eq
which is uniform for $u,$ ${\rm{v}}$ in compact subsets of $(-\infty, 0)$ and $t\in (0,\infty).$
\end{theorem}
\par
The remainder of this paper is organized as follows. In Sect.2, we propose a model Riemann-Hilbert problem associated to $\Phi(\xi, s),$ and derive its Lax pair by proving the Proposition 2. We prove the solvability of the RH problem for $\Phi(\xi, s)$ via a vanishing lemma.
In Sect.3 we apply the Deift-Zhou nonlinear steepest descent method to analyze the RH problem for orthogonal polynomials with respect to the weight funciton (\ref{R1}) and prove Theorem 1. Sect.4 focuses on the reduction of the ${\rm P_{\uppercase\expandafter{\romannumeral5}}}$ kernel to two different Bessel kernels as $s\rw \infty$ and $s\rw 0^{+},$ respectively, and the proof of Theorem 2 and Theorem 3. Sect.5 is to show that the limiting kernel is the Airy kernel at the hard edge and gives the proof of Theorem 4.
\par
We claim that the following Pauli matrices$\sigma_{1},$ $\sigma_{3}$ and two auxiliary matrices $\sigma_{-},$ $\sigma_{+}$ have been used in this paper,
\bbq
\sigma_{1}=\left(
\begin{matrix}
0&1\\
1&0\\
\end{matrix}
\right),
\;\;
\sigma_{3}=\left(
\begin{matrix}
1&0\\
0&-1\\
\end{matrix}
\right),
\;\;
\sigma_{-}=\left(
\begin{matrix}
0&0\\
1&0\\
\end{matrix}
\right),
\;\;
{\rm and}
\;\;
\sigma_{+}=\left(
\begin{matrix}
0&1\\
0&0\\
\end{matrix}
\right).
\eeq
\par
\section{A model RH problem and its solvability}
\par
A model RH problem in the following for $\Phi(\xi, s)$ and it will benefit for the steepest descent analysis.
\par
$(a)$ $\Phi(\xi,s)$ is analytic in $\mathbb{C}\setminus \displaystyle\cup_{j=1}^{3}\widehat{\Sigma}_{j}\cup\left(0,s\right),$ illustrated in Figure 1.
\par
$(b)$ $\Phi(\xi,s)$ fulfills the jump relation
\bq\label{RE20}
\Phi_{+}(\xi,s)=\Phi_{-}(\xi,s)\left\{
\begin{array}{llll}
e^{{i}\lambda\pi{\sigma_{3}}}, & \xi\in \left(0,s\right), \\
\\
\left(
\begin{matrix}
1&0\\
e^{i\pi(\lambda+\alpha)}&1\\
\end{matrix}
\right),& \xi\in \widehat{\Sigma}_{1},\\
\\
\left(
\begin{matrix}
0&1\\
-1&0\\
\end{matrix}
\right),& \xi \in \widehat{\Sigma}_{2},\\
\\
\left(
\begin{matrix}
1&0\\
e^{-i\pi(\lambda+\alpha)}&1\\
\end{matrix}
\right),& \xi \in\widehat{\Sigma}_{3}.
\end{array}
\right.
\eq
\par
$(c)$ For $\xi\rightarrow \infty,$ the asymptotic behavior of $\Phi(\xi,s)$ is given by
\bq\label{RE21}
\Phi(\xi,s)=\left(I+\frac{C_{1}(s)}{\xi}+\mathcal{O}\left(\frac{1}{\xi^{2}}\right)\right)\xi^{-\frac{1}{4}\sigma_{3}}\frac{I+i\sigma_{1}}{\sqrt{2}}e^{\sqrt{\xi}\sigma_{3}},
\eq
where $C_{1}(s)$ only dependents on $s$ and $\arg{\xi} \in (-\pi, \pi).$
\par
$(d)$ For $\xi\rightarrow 0,$ the asymptotic behavior of $\Phi(\xi,s)$ in four sectors $\widehat{\Omega}_{j}, j=1,\ldots,4,$ are given by
\bq\label{RE22}
\Phi(\xi,s)=Q_{1}(s)\left(I+\mathcal{O}(\xi)\right)\xi^{\frac{\alpha}{2}\sigma_{3}}\left\{
\begin{array}{llll}
e^{\frac{{i}\lambda\pi}{2}\sigma_{3}}, & \xi\in \widehat{\Omega}_{1}, \\
\\
\left(
\begin{matrix}
1&0\\
-e^{i\pi(\lambda+\alpha)}&1\\
\end{matrix}
\right),& \xi \in \widehat{\Omega}_{2},\\
\\
\left(
\begin{matrix}
1&0\\
e^{-i\pi(\lambda+\alpha)}&1\\
\end{matrix}
\right),& \xi\in \widehat{\Omega}_{3},\\
\\
e^{-\frac{{i}\lambda\pi}{2}\sigma_{3}}, & \xi\in \widehat{\Omega}_{4}, \\
\end{array}
\right.
\eq
where four sectors $\widehat{\Omega}_{j}, j=1,\ldots,4,$ illustrated in Figure 1, and $Q_{1}(s)$ only depends on $s,$ such that $\det\left(Q_{1}(s)\right)=1.$
\\
$(e)$ For $\xi \rightarrow s,$ the asymptotic behavior $\Phi(\xi,s)$ is given by
\bq\label{RE23}
\Phi(\xi,s)=Q_{2}(s)\left(I+\mathcal{O}(\xi-s)\right)\left(\xi-s\right)^{\frac{\lambda}{2}\sigma_{3}},
\eq
where $Q_{2}(s)$ only depends on $s,$ such that$\det\left(Q_{2}(s)\right)=1,$ and $\arg\left(\xi-s\right) \in (-\pi,\pi).$

\begin{center}
\begin{tikzpicture}
\begin{scope}[line width=2pt]
\draw[->,>=stealth] (0,0)--(1.5,0);
\draw[-] (0,0)--(3,0);
\node[below] at (3,0) {$S$};

\draw[->,>=stealth] (-3.5,0)--(-2,0);
\draw[-] (-4.6,0)--(0,0);
\draw[dashed] (3,0)--(6,0);
\draw[->,>=stealth] (-3.2,2.8)--(-1.6,1.4);
\draw[-] (-2.4,2.1)--(0,0);
\draw[->,>=stealth] (-3.2,-2.8)--(-1.6,-1.4);
\draw[-] (-2.4,-2.1)--(0,0);
\node[below] at (0,0) {$O$};
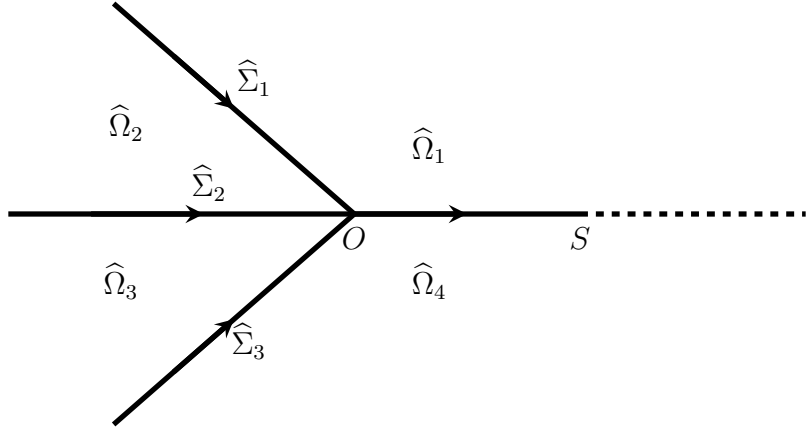
\node[below] at (-.8,-3.5) {Figure 1. Contours $\mathbb{C}\setminus \displaystyle\cup_{j=1}^{3}\widehat{\Sigma}_{j}\cup\left(0,s\right),$ and regions $\widehat{\Omega}_{j}, j=1,\ldots,4.$ };

\node[above] at (-1.4,-2.1) {{$\widehat{\Sigma}_{3}$}};
\node[above] at (-2,0) {{ $\widehat{\Sigma}_{2}$}};
\node[above] at (-1.4,1.4) {{ $\widehat{\Sigma}_{1}$}};

\node[above] at (1,.5) {{$\widehat{\Omega}_{1}$}};
\node[above] at (-3.1,.8) {{ $\widehat{\Omega}_{2}$}};
\node[above] at (-3.1,-1.3) {{$\widehat{\Omega}_{3}$}};
\node[above] at (1,-1.3) {{$\widehat{\Omega}_{4}$}};
\end{scope}
\end{tikzpicture}
\end{center}

The proof of Proposition 2, we follow similar line in \cite{FIK2006} to derive the Lax pair (\ref{RE24}) and (\ref{RE25}) for $\Phi(\xi,s).$
\begin{proof}
From the constant jumps in (\ref{RE20}) of $\Phi(\xi,s),$ therefore $\det\left(\Phi(\xi,s)\right)$ is an entire function.
From (\ref{RE21}), $\det\left(\Phi(\xi,s)\right)=1+\mathcal{O}(1/\xi)$ is uniform for $\xi \rightarrow \infty$ in $\mathbb{C}\setminus \displaystyle\cup_{j=1}^{3}\widehat{\Sigma}_{j}\cup\left(0,s\right).$
Hence, by Liouville's theorem, one finds $\det\left(\Phi(\xi,s)\right)=1.$ So, ${\rm tr}C_{1}(s)=0$ is valid. We suppose
\bq\label{RE34}
C_{1}(s)=\left(
\begin{matrix}
q(s)&-ir(s)\\
it(s)&-q(s)\\
\end{matrix}
\right).
\eq
\par
Both rational matrix functions $\Phi_{s}\Phi^{-1}$ and $\Phi_{\xi}\Phi^{-1}$ are analytic in $\xi$ plane, with two possible simple poles $0$ and $s,$
since all the jumps in (\ref{RE20}) of the RH problem for $\Phi(\xi,s)$ are constant matrices.
\par
The rational function $\Phi_{\xi}\Phi^{-1}$ has a removable singularity at $\xi=\infty,$ and two possible simple poles at $\xi=0,$ $\xi=s,$ respectively, by the asymptotic behaviors (\ref{RE21}), (\ref{RE22}) and (\ref{RE23}) of $\Phi(\xi,s).$ This leads to
\bq\label{RE35}
\Phi_{\xi}\Phi^{-1}=A_{0}(s)+\frac{A_{1}(s)}{\xi}+\frac{A_{2}(s)}{\xi-s}.
\eq
\par
The asymptotic behavior of the LHS of (\ref{RE35}) at $\xi=\infty$ derives from (\ref{RE21}) and expanding the RHS of (\ref{RE35}) at $\xi=\infty.$ Comparing the constant terms and the coefficients of $\xi^{-1}$ on both sides, then one finds,
\bq\label{RE36}
A_{0}=\frac{i}{2}\sigma_{-},\;\;A_{1}+A_{2}=-\frac{1}{4}\sigma_{3}+\frac{i}{2}\left(C_{1}\sigma_{-}-\sigma_{-}C_{1}\right)-\frac{i}{2}\sigma_{+}.
\eq
\par
A calculation of the LHS of (\ref{RE35}) with (\ref{RE22}) and comparing the coefficients of $\xi^{-1}$ on both sides as $\xi\rw 0,$ one has,
\bq\label{RE37}
A_{1}=\frac{\alpha}{2}Q_{1}(s)\sigma_{3}Q_{1}^{-1}(s), \;\;\; \det(A_{1})=-\frac{\alpha^{2}}{4}.
\eq
\par
Similarly, a calculation of the LHS of (\ref{RE35}) with (\ref{RE23}) and comparing the coefficients of $(\xi-s)^{-1}$ on both sides as $\xi\rw s,$ one finds,
\bq\label{RE38}
A_{2}=\frac{\lambda}{2}Q_{2}(s)\sigma_{3}Q_{2}^{-1}(s), \;\;\; \det(A_{2})=-\frac{\lambda^{2}}{4}.
\eq
\par
Moreover, the rational function $\Phi_{s}\Phi^{-1}$ has a removable singularity at $\xi=\infty,$ and a simple pole at $\xi=s.$ From the asymptotic behaviors (\ref{RE21}) and (\ref{RE23}) of $\Phi(\xi,s),$ this provides
\bq\label{RE39}
\Phi_{s}\Phi^{-1}=B_{0}(s)+\frac{B_{2}(s)}{\xi-s}.
\eq
\par
Similarly, substituting (\ref{RE23}) into the rational function $\Phi_{\xi}\Phi^{-1},$ and comparing the coefficient of $(\xi-s)^{-1}$ on both sides of (\ref{RE39}) as $\xi\rw s$, then one finds,
\bq\label{RE40}
B_{2}=-\frac{\lambda}{2}Q_{2}(s)\sigma_{3}Q_{2}^{-1}(s), \;\;\; \det(B_{2})=-\frac{\lambda^{2}}{4}.
\eq
\par
Inserting (\ref{RE21}) into the rational function $\Phi_{\xi}\Phi^{-1}$, and comparing the constant terms and the coefficients of $\xi^{-1}$ on both sides of (\ref{RE39}), one has, $B_{0}(s)=0$ and $B_{2}(s)=C_{1}'(s),$ furthermore, with the equations (\ref{RE34}), (\ref{RE36}), (\ref{RE38}) and (\ref{RE40}), then (\ref{RE24}), (\ref{RE25}), (\ref{RE26}) and (\ref{RE27}) are valid.
\par
From the mixed derivatives $\Phi_{\xi{s}}=\Phi_{s\xi},$ which leads to the compatibility condition
\bq\label{RE41}
A_{s}-B_{\xi}=BA-AB,
\eq
where
\bbq
A=A_{0}(s)+\frac{A_{1}(s)}{\xi}+\frac{A_{2}(s)}{\xi-s},\quad B=\frac{B_{2}(s)}{\xi-s}.
\eeq
After some calculation and simplify, the compatibility condition (\ref{RE41}) leads to
\bq\label{RE42}
sA_{1}'(s)=A_{1}(s)B_{2}(s)-B_{2}(s)A_{1}(s),
\eq
where $A_{1}(s),$ $B_{2}(s)$ are in (\ref{RE26}) and (\ref{RE27}).
Rewriting the equation (\ref{RE42}), $\det(A_{1})=-\frac{\alpha^{2}}{4}$ (see (\ref{RE37})) and $\det(B_{2})=-\frac{\lambda^{2}}{4}$ (see (\ref{RE40})) in terms of the auxiliary functions $q(s), r(s), t(s)$ and their derivatives, after some calculations, one gets,
\bq\label{RE43}
q''(s)=\frac{t'(s)}{2s}+\frac{r'(s)q(s)}{s}-\frac{r'(s)}{2},
\eq

\bq\label{RE44}
r''(s)=-\frac{r'(s)}{2s}+\frac{r'(s)r(s)}{s}-\frac{q'(s)}{s},
\eq

\bq\label{RE45}
q'(s)^{2}+r'(s)t'(s)=\frac{\lambda^{2}}{4},
\eq

\bq\label{RE46}
\left(q'(s)+\frac{1}{2}r(s)-\frac{1}{4}\right)^{2}+\left(\frac{1}{2}+r'(s)\right)\left(t'(s)-q(s)\right)=\frac{\alpha^{2}}{4}.
\eq
\par
Although the above system has four equations with three unknown functions, it is solvable. In fact, the first equation is equivalent to a suitable combination of the rest three equations (\ref{RE44})-(\ref{RE46}).
With the aid of (\ref{RE45}), the first equation (\ref{RE43}) is equivalent to
\bq\label{RRE01}
q''(s)=\frac{\lambda^{2}-4q'(s)^{2}}{8sr'(s)}-\frac{r'(s)}{2}+\frac{q(s)r'(s)}{s}.
\eq
Substituting (\ref{RE44}) and (\ref{RE45}) into the derivative of the fourth equation (\ref{RE46}), then one finds the following equation,
\bq\label{RRE02}
\frac{r''(s)}{r'(s)}\left(q''(s)-\frac{\lambda^{2}-4q'(s)^{2}}{8sr'(s)}+\frac{r'(s)}{2}-\frac{q(s)r'(s)}{s}\right)=0,
\eq
which is equivalent to the equation (\ref{RRE01}), so the first equation (\ref{RE43}) in the system can be neglected. From (\ref{RE44}), (\ref{RE45}) and (\ref{RE46}), one finds that $q'(s),$ $t'(s)$ and $q(s)$ in terms of $r(s)$ and its derivatives, these are explicit in (\ref{RE28}), (\ref{RE29}) and (\ref{RE30}).
\end{proof}
\par
{\bfseries Solvability of the Riemann-Hilbert problem for $\Phi(\xi,s)$}
\par
In general case, the solvability of a RH problem derives from the triviality of its homogeneous RH problem, namely the vanishing lemma. With the aid of the Cauchy operator, a RH problem turns out to be a Fredholm singular equation, for examples \cite{ DeiftBook, DKV1999, FZ1992, FIK2006, IKO2008}. The vanishing lemma follows that the null sapce is trival,  and with the Fredholm alternative theorem, imply that the index of the Fredholm singular equation is zero and the RH problem is solvable. A summary contained in \cite{FIK2006} and references therein. Some examples and details see also \cite{CIK2011, KMM2003, XZ2015}. Here, a vanishing lemma proved for the homogeneous RH problem $\widetilde{\Phi}(\xi, s).$

\begin{lemma}
Suppose that $\widetilde{\Phi}(\xi,s)$ satisfies the jump conditions (\ref{RE20}) and the boundary conditions (\ref{RE22}), (\ref{RE23}) of the RH problem for $\Phi(\xi,s),$ with the asymptotic behavior (\ref{RE21}) at infinity replaced by
\bq\label{RE50}
\widetilde{\Phi}(\xi,s)=\mathcal{O}\left(\frac{1}{\xi}\right)\xi^{-\frac{1}{4}\sigma_{3}}\frac{I+i\sigma_{1}}{\sqrt{2}}e^{\sqrt{\xi}\sigma_{3}}, \;\; {\rm as} \;\; \xi\rightarrow\infty.
\eq
If $s \in (0, \infty),$ then it follows that the solution is trivial, $\widetilde{\Phi}(\xi,s)\equiv0.$
\end{lemma}

\begin{proof}
In order to eliminate the exponential factor at infinity and remove the jumps on $\widehat{\Sigma}_{1}$ and $\widehat{\Sigma}_{3},$ let $\widetilde{\Phi}_{1}(\xi,s)$ be defined as follows
\bq\label{RE51}
\widetilde{\Phi}_{1}(\xi,s)=\left\{
\begin{array}{llll}
\widetilde{\Phi}(\xi,s)e^{-\sqrt{\xi}\sigma_{3}}, & \xi \in \widehat{\Omega}_{1}\cup\widehat{\Omega}_{4},\\
\\
\widetilde{\Phi}(\xi,s)e^{-\sqrt{\xi}\sigma_{3}}\left(
\begin{matrix}
1&0\\
e^{-2\sqrt{\xi}}e^{i\pi(\lambda+\alpha)}&1\\
\end{matrix}
\right),& \xi \in \widehat{\Omega}_{2},\\
\\
\widetilde{\Phi}(\xi, s)e^{-\sqrt{\xi}\sigma_{3}}\left(
\begin{matrix}
1&0\\
-e^{-2\sqrt{\xi}}e^{-i\pi(\lambda+\alpha)}&1\\
\end{matrix}
\right),& \xi\in \widehat{\Omega}_{3},\\
\end{array}
\right.
\eq
where the sectors $\widehat{\Omega}_{j}, j=1,\ldots,4,$ illustrated in Figure 1, $\arg\xi \in (-\pi, \pi).$
\par
Then $\widetilde{\Phi}_{1}(\xi, s)$ fulfills the following RH conditions:\\
$(a)$ $\widetilde{\Phi}_{1}(\xi, s)$ is analytic in $\xi \in \mathbb{C} \setminus \widehat{\Sigma}_{2}\cup (0,s),$
\\
$(b)$ $\widetilde{\Phi}_{1}(\xi, s)$ satisfies the following jump condition
\bq\label{RE52}
(\widetilde{\Phi}_{1})_{+}(\xi, s)=(\widetilde{\Phi}_{1})_{-}(\xi, s)\left\{
\begin{array}{ll}
e^{i\lambda{\pi}\sigma_{3}}, & \xi \in (0, s),\\
\\
\left(
\begin{matrix}
e^{i\pi(\alpha+\lambda)}e^{-2\sqrt{\xi_{+}}}&1\\
0&e^{-i\pi(\alpha+\lambda)}e^{2\sqrt{\xi_{+}}}\\
\end{matrix}
\right),& \xi \in \widehat{\Sigma}_{2},\\
\end{array}
\right.
\eq
where $\sqrt{\xi_{+}}=i\sqrt{|\xi|}$ for $\xi\in \widehat{\Sigma}_{2}\cup (0, s).$
\\
$(c)$ The asymptotic behavior of $\widetilde{\Phi}_{1}(\xi,s)$ at $\xi=\infty$ is
\bq\label{RE53}
\widetilde{\Phi}_{1}(\xi,s)=\mathcal{O}\left(\xi^{-\frac{3}{4}}\right), \;\;\arg\xi \in (-\pi,\pi).
\eq
\\
$(d)$ The asymptotic behavior of  $\widetilde{\Phi}_{1}(\xi,s)$ at $\xi=0$ is
\bq\label{RE54}
\widetilde{\Phi}_{1}(\xi,s)=\mathcal{O}(1)\left(\xi-s\right)^{\frac{\lambda}{2}\sigma_{3}}\xi^{\frac{\alpha}{2}\sigma_{3}}.
\eq
\\
$(e)$ The asymptotic behavior of $\widetilde{\Phi}_{1}(\xi,s)$ at $\xi=s$ is
\bbq\label{RE55}
\widetilde{\Phi}_{1}(\xi,s)=\mathcal{O}(1)\left(\xi-s\right)^{\frac{\lambda}{2}\sigma_{3}}.
\eeq
\par
In order to translate the oscillation terms in diagonal to off-diagonal, another transformation applied in the following,
\bq\label{RE56}
\widetilde{\Phi}_{2}(\xi,s)=\left\{
\begin{array}{ll}
\widetilde{\Phi}_{1}(\xi,s)\left(
\begin{matrix}
0&-1\\
1&0\\
\end{matrix}
\right),& {\rm Im}\xi>0,\\
\\
\widetilde{\Phi}_{1}(\xi,s), & {\rm Im}\xi<0.\\
\end{array}
\right.
\eq
\par
Then $\widetilde{\Phi}_{2}(\xi,s)$ satisfies the RH problem as follows
\par
$(a)$ $\widetilde{\Phi}_{2}(\xi,s)$ is analytic in $\mathbb{C} \setminus \mathbb{R}.$
\par
$(b)$ $\widetilde{\Phi}_{2}(\xi,s)$ satisfies the following jump condition
\\
\bq\label{R57}
(\widetilde{\Phi}_{2})_{+}(\xi,s)=(\widetilde{\Phi}_{2})_{-}(\xi,s)\left\{
\begin{array}{llll}
\left(
\begin{matrix}
1&-e^{-2\sqrt{\xi_{+}}}e^{{i}(\alpha+\lambda)\pi}\\
e^{2\sqrt{\xi_{+}}}e^{-{i}(\alpha+\lambda)\pi}&0\\
\end{matrix}
\right),& \xi \in (-\infty, 0),\\
\\
\left(
\begin{matrix}
0&-e^{{i}\lambda\pi}\\
e^{-{i}\lambda\pi}&0\\
\end{matrix}
\right),& \xi\in (0, s),\\
\\
\left(
\begin{matrix}
0&-1\\
1&0\\
\end{matrix}
\right),& \xi\in (s, +\infty).\\
\end{array}
\right.
\eq
\par
$(c)$ The behavior of $\widetilde{\Phi}_{2}(\xi,s)$ at infinity is the same as (\ref{RE53}).
\par
$(d)$ The behavior of $\widetilde{\Phi}_{2}(\xi,s)$ at $\xi=0$ is
\par
\bq\label{R58}
\widetilde{\Phi}_{2}(\xi,s)=\mathcal{O}(1)\left(\xi-s\right)^{\frac{\lambda}{2}\sigma_{3}}\xi^{\frac{\alpha}{2}\sigma_{3}}\left\{
\begin{array}{ll}
\left(
\begin{matrix}
0&-1\\
1&0\\
\end{matrix}
\right),& {\rm Im}\xi>0,\\
\\
I, & {\rm Im}\xi<0.\\
\end{array}
\right.
\eq
\par
$(e)$ The behavior of $\widetilde{\Phi}_{2}(\xi,s)$ at $\xi=s,$
\\
\bq\label{R59}
\widetilde{\Phi}_{2}(\xi,s)=\mathcal{O}(1)\left(\xi-s\right)^{\frac{\lambda}{2}\sigma_{3}}\left\{
\begin{array}{ll}
\left(
\begin{matrix}
0&-1\\
1&0\\
\end{matrix}
\right),& {\rm Im}\xi>0,\\
\\
I, & {\rm Im}\xi<0.\\
\end{array}
\right.
\eq
For later convenience, an auxiliary matrix function defined as follows
\bbq\label{R60}
H(\xi)=\widetilde{\Phi}_{2}(\xi,s)\left(\widetilde{\Phi}_{2}(\overline{\xi},s)\right)^{*}, \;\; \xi \notin \mathbb{R},
\eeq
where $Z^{*}$ denotes the Hermitian conjugate of $Z.$ With the asymptotic behavior of $\widetilde{\Phi}_{2}(\xi,s)$ at $\xi=\infty$ (\ref{RE53}), $\xi=0$ (\ref{R58}), and $\xi=s$ (\ref{R59}), one finds,
\bbq\label{}
H(\xi)=\mathcal{O}\left(\xi^{-\frac{3}{2}}\right),\; \xi\rightarrow \infty;\;\;\; H(\xi)=\mathcal{O}(1),\; \xi\rightarrow 0;\;\;\; H(\xi)=\mathcal{O}(1),\; \xi\rightarrow s.
\eeq
Then, $H_{+}(\xi)$ is integrable over the real axis, by Cauchy's formula, one has
\bq\label{R61}
\int_{R}H_{+}(\xi)d\xi=0.
\eq
\par
From (\ref{R57}), the sum of equation (\ref{R61}) and its Hermitian conjugate, then one finds,
\bbq\label{R62}
\int_{-\infty}^{0}(\widetilde{\Phi}_{2})_{-}(\xi,s)\left(
\begin{matrix}
2&0\\
0&0\\
\end{matrix}
\right)\left((\widetilde{\Phi}_{2})_{-}\right)^{*}(\xi,s)d\xi
=
\int_{-\infty}^{0}\left(
\begin{matrix}
|((\widetilde{\Phi}_{2})_{11})_{-}|^{2}&\bullet\\
\bullet&|((\widetilde{\Phi}_{2})_{21})_{-}|^{2}\\
\end{matrix}\right)d\xi=0.
\eeq
From the above equation, for $\xi\in (-\infty, 0),$ the first column of $(\widetilde{\Phi}_{2})_{-}(\xi,s)$ vanishes. By the jump condition (\ref{R57}) on the interval $(-\infty, 0),$ the second column of $(\widetilde{\Phi}_{2})_{+}(\xi,s)$ also vanishes.
\par
With the aid of the theorem due to Carlson in \cite{RS1987}, we prove the rest entries of $\widetilde{\Phi}_{2}(\xi,s)$ also vanish. We follow similar line in \cite{XDZ2014}, to construct a auxiliary function $G_{j}(\xi), j=1, 2,$ in the following,
\bbq\label{RE63}
G_{j}(\xi)=\left\{
\begin{array}{ll}
\left(\widetilde{\Phi}_{2}(\xi,s)\right)_{j1},& {\rm Im}\xi>0,\\
\\
\left(\widetilde{\Phi}_{2}(\xi,s)\right)_{j2}, & {\rm Im}\xi<0,\\
\end{array}
\right.
\eeq
then $G_{j}$ is analytic in $\mathbb{C}\setminus (-\infty, s).$ By (\ref{R57}), then $G_{j}(\xi) (j=1,2)$ satisfy the jump condition
\bbq\label{RE64}
G_{j+}(\xi)=G_{j-}(\xi)\left\{
\begin{array}{ll}
e^{2\sqrt{\xi_{+}}-{i}(\alpha+\lambda)\pi},& \xi\in (-\infty, 0),\\
\\
e^{-{i}\lambda\pi},\;\; &\xi\in (0, s).\\
\end{array}
\right.
\eeq
\par
We extend scalar function $G_{j}(\xi)$ in the following for $\xi\in (-\infty, 0),$
\bbq\label{RE65}
\widehat{G}_{j}(\xi)=\left\{
\begin{array}{ll}
G_{j}\left(e^{-{i}2\pi}\xi\right)e^{2\sqrt{\xi}}e^{-i(\alpha+\lambda)\pi}, & \pi\leq \arg\xi <2\pi,\\
\\
G_{j}\left(e^{{i}2\pi}\xi\right)e^{2\sqrt{\xi}}e^{i(\alpha+\lambda)\pi}, & -2\pi<\arg\xi\leq-\pi,\\
\end{array}
\right.
\eeq
then $\widehat{G}_{j}(\xi)$ is analytic in a large sector $-2\pi<\arg \xi<2\pi$ and $\xi\notin[0,s].$
\par
Let
\bbq\label{RE66}
h_{j}(\xi)=\widehat{G}_{j}\left((\xi+s+1)^{4}\right),\;\; {\rm for}\;\;\; {\rm Re}{\xi}\geq 0,
\eeq
hence $h_{k}(\xi)$ is analytic in ${\rm Re}\xi>0,$ continuous and bounded in ${\rm Re}\xi\geq 0,$ and fulfills
\bbq\label{RE67}
\left|h_{j}(\xi)\right|=\mathcal{O}\left(e^{-\left|\xi\right|^{2}}\right),\;\; {\rm for}\; \arg\xi=\pm\frac{\pi}{2}\; {\rm and}\; \left|\xi\right|\rightarrow \infty,
\eeq
by Carlson's theorem, one gets $h_{j}(\xi)\equiv 0,$ $j=1,2$ for ${\rm Re}\xi\geq 0.$ Tracing back steps, $(\widetilde{\Phi}_{2})_{+}(\xi, s)$ vanishes identically for $\xi\in \mathbb{C}\setminus \mathbb{R}.$ Then, $\widetilde{\Phi}(\xi, s)\equiv 0$ by (\ref{RE51}) and (\ref{RE56}). This proves the vanishing lemma.
\end{proof}

\begin{rem}
Note that the ``free'' parameter $\lambda$ is valid for $s>0,$ and $\al>0$ is also necessary. One checks by substituting (\ref{RE54}) into (\ref{RE52}) for $\xi \in \widehat{\Sigma}_{2},$ and $\xi\rightarrow 0,$ then one has
\bq\label{R18}
\mathcal{O}(1)=\left(
\begin{matrix}
e^{-2\sqrt{\xi_{+}}}&\left(|\xi|+s\right)^{\lambda}|\xi|^{\alpha}\\
0&e^{2\sqrt{\xi_{+}}}\\
\end{matrix}
\right),
\eq
where $\mathcal{O}(1)$ is a bounded, and with determinant $1.$ From factors $\left(|\xi|+s\right)^{\lambda}$ and $|\xi|^{\alpha}$ in the above matrix, the solvability of $\Phi(\xi,s)$ confined for $s>0$ and $\al>0.$
\end{rem}

\section{Deift-Zhou steepest descent analysis}
\par
With the well-known relation presented in \cite{FIK1992}, the orthogonal polynomials with respect to the weight (\ref{R1}) can be characterized by a RH problem for $Y.$ We adopt the powerful Deift-Zhou steepest descent analysis method (or Riemann-Hilbert method) in \cite{DZ1993}, see also \cite{BI1999, DX1999, DKV1999}, to analyze the RH problem for $Y.$ Following the standard process, we obtain a series of invertible transformations $Y\rw T\rw S\rw R,$ at last, the matrix function $R$ is close to the identity matrix. After that, taking a list of inverse transformations, then the kernel (\ref{00a1}) associated with the weight (\ref{R1})
can be represented by the uniform asymptotics of the orthogonal polynomials in the complex plane for large $n.$
\subsection{Riemann-Hilbert problem for $Y$}
\par
The orthogonal polynomials with respect to the weight function $w(x, t)$ in (\ref{R1}) are described by the following $2\times2$ matrix valued function $Y(z).$
\par
$(a)$ $Y(z)$ is analytic for $z\in \mathbb{C}\setminus [0, \infty).$
\par
$(b)$ $Y(z)$ satisfies the jump condition
\begin{equation*}\label{RE1}
Y_{+}(x)=Y_{-}(x)\left(
\begin{matrix}
1&w(x,t)\\
0&1\\
\end{matrix}
\right),\quad {\rm for}\quad x\in (0,+\infty),
\end{equation*}
where $w(x, t)$ is given by (\ref{R1}).
\par
$(c)$ The asymptotic behavior of $Y(z)$ at infinity is
\begin{equation*}\label{RE2}
Y(z)=\left(I+\mathcal{O}\left(z^{-1}\right)\right)\left(
\begin{matrix}
z^{n}&0\\
0&z^{-n}\\
\end{matrix}
\right),\quad z\rightarrow\infty.
\end{equation*}
\par
$(d)$ The asymptotic behavior of $Y(z)$ at the origin is
\begin{equation*}\label{RE3}
Y(z)=\left(
\begin{matrix}
O(1)&O(1)\\
O(1)&O(1)\\
\end{matrix}
\right)
,\quad z\rightarrow 0.
\end{equation*}
\par
By the work of Fokas, Its and Kitaev \cite{FIK1992}, the above RH problem for $Y(z)$ has a unique solution,
\begin{equation}\label{RE4}
Y(z)=\left(
\begin{matrix}
\pi_{n}(z)&\frac{1}{2\pi{i}}\int_{0}^{\infty}\frac{\pi_{n}(s)w(s)}{s-z}ds\\
-2\pi{i}\gamma_{n-1}^{2}\pi_{n-1}(z)&-\gamma_{n-1}^{2}\int_{0}^{\infty}\frac{\pi_{n-1}(s)w(s)}{s-z}ds\\
\end{matrix}
\right),
\end{equation}
where $\pi_{n}(z)$ is the monic polynomial, and $P_{n}(z)=\gamma_{n}\pi_{n}(z)$ is the orthonormal polynomial with respect to the weight $w(x,t),$ see (\ref{R1}).
\par
\subsection{Riemann-Hilbert problem for $T$}
\par
In order to normalize the matrix function $Y(z)$ at infinity, we rescale the variable and introduce the first transformation $Y\rightarrow T,$ and $T$ is defined as follows
\begin{equation}\label{RW2}
T(z)=(4n)^{-(n+\frac{\alpha+\lambda}{2})\sigma_{3}}e^{-\frac{n\ell}{2}\sigma_{3}}Y(4nz)e^{-n(g(z)-\frac{\ell}{2})\sigma_{3}}
\left(4n\right)^{\frac{\alpha+\lambda}{2}\sigma_{3}},
\end{equation}
for $z\in \mathbb{C}\setminus [0,+\infty),$ where $\ell=-2(1+2\ln{4})$ is the Euler-Laguerre constant.
\par
Then $T(z)$ solves the following RH problem,
\par
$(a)$ $T(z)$ is analytic for $z\in \mathbb{C}\setminus [0, \infty).$
\par
$(b)$ $T(z)$ satisfies the following jump condition,
\begin{equation}\label{RE5}
T_{+}(x)=T_{-}(x)
\begin{array}{ll}
\left(
\begin{matrix}
e^{n(g_{-}(x)-g_{+}(x))}&x^{\alpha}\left(x+\frac{t}{4n}\right)^{\lambda}e^{n(g_{-}(x)+g_{+}(x)-4x-\ell)}\\
0&e^{n(g_{+}(x)-g_{-}(x))}\\
\end{matrix}
\right),\;\; x\in (0,+\infty).
\end{array}
\end{equation}
\par
$(c)$ The asymptotic behavior of $T(z)$ at infinity is
\begin{equation}\label{T25}
T(z)=I+\mathcal{O}\left(z^{-1}\right),\quad z\rightarrow\infty.
\end{equation}
\par
$(d)$ The asymptotic behavior of $T(z)$ at the origin is
\begin{equation}\label{RE6}
T(z)=\left(
\begin{matrix}
O(1)&O(1)\\
O(1)&O(1)\\
\end{matrix}
\right)
, \quad z \rightarrow 0.
\end{equation}
\par
For the classical Laguerre weight $x^{\al}e^{-4x},$ $x>0,$ $\al>-1,$ the equilibrium measure is given by $\mu(x)=\frac{2}{\pi}\sqrt{\frac{1-x}{x}},$ $0<x<1,$ see (\ref{R24}), which is independent of $\al,$
see \cite{V2007, QW2008}. From (\ref{RE5}), we can construct $g(z)$ with the equilibrium measure of the classical Laguerre weight for large $n$ and achieve the normalization of $Y(z)$ at infinity. Similar method see \cite{XDZ2014}. We define several auxiliary functions and refer \cite{QW2008},
\bq\label{T16}
g(z)=\int_{0}^{1}\log(z-x)\mu(x)dx,
\eq
where $\arg(z-x)\in (-\pi,\pi),$ and
\bq\label{RW01}
\phi(z)=2\int_{0}^{z}\sqrt{\frac{s-1}{s}}ds,\quad z\in \C \setminus [0,+\infty),
\eq
where $\arg{z}\in (0,2\pi).$
\par
From the definition of $g(z)$ and $\phi(z),$ the following properties are easily to check,
\bq\label{T23}
g_{+}(x)=g_{-}(x),\quad x\in (1,+\infty),
\eq
\bq\label{T24}
g_{+}(x)+g_{-}(x)-4x-\ell=-2\phi(x), \quad x\in (1+\infty),
\eq
and
\bq\label{a4}
g_{+}(x)-g_{-}(x)=2\pi{i}-2\phi_{+}(x)=2\pi{i}+2\phi_{-}(x), x \in (0,1),
\eq
\bq\label{a5}
g_{+}(x)+g_{-}(x)-4x-\ell=0, \quad x \in (0,1),
\eq
where $\ell$ is the Euler-Lagrange constant. With the aid of the above properties of $g(z)$ and $\phi(z),$ see (\ref{T23}), (\ref{T24}), (\ref{a4}) and (\ref{a5}), then the jump conditions for $T$ in (\ref{RE5}) can be expressed in terms of $\phi$ and simplify as follows
\bq\label{RE5}
T_{+}(x)=T_{-}(x)\left\{
\begin{array}{ll}
\left(
\begin{matrix}
e^{2n\phi_{+}(x)}&x^{\alpha}\left(x+\frac{t}{4n}\right)^{\lambda}\\
0&e^{2n\phi_{-}(x)}\\
\end{matrix}
\right), & x\in (0,1),
\\
\left(
\begin{matrix}
1&x^{\alpha}\left(x+\frac{t}{4n}\right)^{\lambda}e^{-2n\phi(x)}\\
0&1\\
\end{matrix}
\right),& x\in \left(1,+\infty\right).
\end{array}
\right.
\eq
\par
In order to remove the oscillation diagonal entries in the above jump matrix for $x \in (0,1),$ the contour can be deformed as an opening lens with the matrix factorization. We define the following piecewise analytic function $S(z).$
\par
\subsection{Riemann-Hilbert problem for $S$}
We introduce the second transformation $T\rightarrow S,$ and $S(z)$ is defined as
\begin{equation}\label{RE13}
S(z)=\left\{
\begin{array}{lll}
T(z),&{\rm for}\; z\; {\rm outside}\; {\rm the}\; {\rm lens}\; {\rm shaped}\; {\rm region},\\
\\
T(z)\left(
\begin{matrix}
1&0\\
-z^{-\alpha}\left(z+\frac{t}{4n}\right)^{-\lambda}e^{2n\phi(z)}&1\\
\end{matrix}
\right),&{\rm for}\; z\; {\rm in}\; {\rm the}\; {\rm upper}\; {\rm lens}\; {\rm region},\\
\\
T(z)\left(
\begin{matrix}
1&0\\
z^{-\alpha}\left(z+\frac{t}{4n}\right)^{-\lambda}e^{2n\phi(z)}&1\\
\end{matrix}
\right),& {\rm for}\; z\; {\rm in}\; {\rm the}\; {\rm lower}\; {\rm lens}\; {\rm region},
\end{array}
\right.
\end{equation}
where $\arg z \in (-\pi,\pi).$
\par
Combining the conditions (\ref{RE5}), (\ref{T25}) and (\ref{RE6}) of the RH problem for $T$ and the definition (\ref{RE13}) of $S,$ then $S$ satisfies the following RH problem,
\par
$(a)$ $S(z)$ is analytic in $\C\setminus \{\bigcup_{k=1}^{3}\Sigma_{k}\}\cup\left(1, \infty\right),$ illustrated in Fig.2.
\par
$(b)$ $S_{+}(z)=S_{-}(z)J_{S}$ for $z \in \{\cup_{k=1}^{3}\Sigma_{k}\}\cup\left(1, \infty\right),$ where the jump $J_{S}$ is
\bq\label{RE14}
J_{S}(z)=\left\{
\begin{array}{lll}
\left(
\begin{matrix}
1&0\\
z^{-\alpha}\left(z+\frac{t}{4n}\right)^{-\lambda}e^{2n\phi(z)}&1\\
\end{matrix}
\right),&{\rm for}\; z \in \Sigma_{1}\bigcup\Sigma_{3},\\
\\
\left(
\begin{matrix}
0&x^{\alpha}\left(x+\frac{t}{4n}\right)^{\lambda}\\
-x^{-\alpha}\left(x+\frac{t}{4n}\right)^{-\lambda}&0\\
\end{matrix}
\right),& {\rm for}\;z=x\in \Sigma_{2}=(0,1),\\
\\
\left(
\begin{matrix}
1&x^{\alpha}\left(x+\frac{t}{4n}\right)^{\lambda}e^{2n\phi(x)}\\
0&1\\
\end{matrix}
\right),& {\rm for}\; z=x\in (1,+\infty).
\end{array}
\right.
\eq
\par
$(c)$ The asymptotic behavior at infinity is
\bbq
S(z)=I+\mathcal{O}\left(z^{-1}\right), \;{\rm as}\;\; z\rightarrow \infty.
\eeq
\par
$(d)$ The asymptotic behavior at the origin is
\begin{equation}\label{RE15}
S(z)=\mathcal{O}(1)\left\{
\begin{array}{lll}
I,&{\rm outside}\; {\rm the}\; {\rm lens}\; {\rm region},\\
\\
\left(
\begin{matrix}
1&0\\
-z^{-\alpha}\left(z+\frac{t}{4n}\right)^{-\lambda}e^{2n\phi(z)}&1\\
\end{matrix}
\right),&{\rm in}\; {\rm the}\; {\rm upper}\; {\rm lens}\; {\rm region},\\
\\
\left(
\begin{matrix}
1&0\\
z^{-\alpha}\left(z+\frac{t}{4n}\right)^{-\lambda}e^{2n\phi(z)}&1\\
\end{matrix}
\right),& {\rm in}\; {\rm the}\; {\rm lower}\; {\rm lens}\; {\rm region}.
\end{array}
\right.
\end{equation}

\begin{center}
\begin{tikzpicture}
\begin{scope}[line width=2pt]
\draw[->,>=stealth] (0,0)--(3,0);

\draw[-] (0,0)--(10,0);
\draw[->,>=stealth] (6,0)--(8,0);
\draw[->,>=stealth] (2.9,2.25)--(3.1,2.25);
\draw[->,>=stealth] (2.9,-2.25)--(3.1,-2.25);

\draw[-] (3,0)--(5,0);
\draw (0,0) .. controls (2.9,3) and (3.1,3) .. (6,0);
\draw (0,0) .. controls (2.9,-3) and (3.1,-3) .. (6,0);
\node[below] at (0,0) {$O$};
\node[above] at (1.8,2) {$\Sigma_{3}$};
\node[above] at (1.8,0) {$\Sigma_{2}$};
\node[above] at (1.8,-1.5) {$\Sigma_{1}$};
\node[below] at (6,0) {$1$};
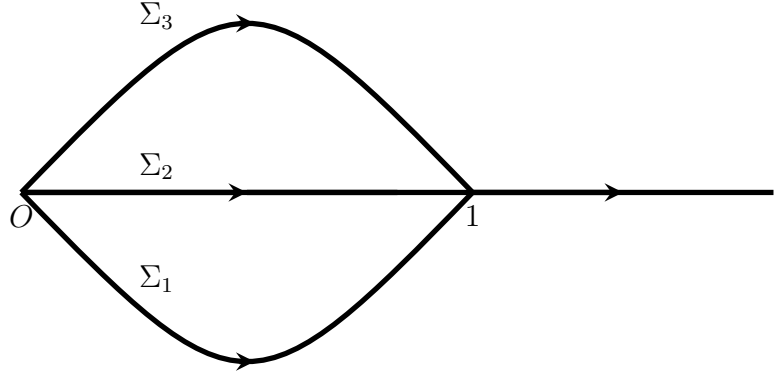
\node[below] at (3.5,-2.9) { Figure 2. The contour $\{\bigcup_{k=1}^{3}\Sigma_{k}\}\bigcup(1,\infty)$ for $S(z).$};
\end{scope}
\end{tikzpicture}
\end{center}

\subsection{Global parametrix}
\par
For $n\rw \infty,$ due to the exponentially small term in (\ref{RE14}), the jump matrix $J_{S}$ on $\Sigma_{1},$ $\Sigma_{3}$ and $(1,\infty)$ converges to the identity matrix very quickly. Moreover, $J_{S}$ on $\Sigma_{2},$ the items $\left(x+t/4n\right)^{\pm\lambda}=x^{\pm\lambda}+\mathcal{O}(n^{-1}),$ $x>0,$ and $t\in (0,c],$ $0<c<\infty,$ as $n\rw \infty.$ Then $S(z)$ can be approximated by a solution of the following RH problem for $P^{(\infty)}(z).$
\par
$(a)$ $P^{(\infty)}(z)$ is analytic in $\C \setminus [0,1].$
\par
$(b)$ $P^{(\infty)}(z)$ satisfies the jump condition,
\bq\label{RE16}
P^{(\infty)}_{+}(x)=P^{(\infty)}_{-}(x)
\left(
\begin{matrix}
0&x^{\al+\la}\\
-x^{-(\al+\la)}&0\\
\end{matrix}
\right),\;\; {\rm for}\;\;x \in \Sigma_{2}=(0,1).\\
\eq
\par
$(c)$ The asymptotic behavior at infinity is
\bbq
P^{(\infty)}(z)=I+\mathcal{O}\left(z^{-1}\right),\;\; z\rightarrow \infty.
\eeq
\par

In the spirit of \cite{DeiftBook} and \cite{KM2004}, the unique solution to the above RH problem can be constructed as follows
\bq\label{R14}
P^{(\infty)}(z)=D(\infty)^{\sigma_{3}}M^{-1}a^{-\sigma_{3}}(z)MD^{\sigma_{3}}(z), \;\; {\rm for}\;\; z\in \mathbb{C}\setminus [0,1],
\eq
where $M=(I+i\sigma_{1})/\sqrt{2},$ $a(z)=((z-1)/z)^{\frac{1}{4}},$ $D(\infty)=2^{-(\al+\la)},$ $D(z)=(\sqrt{z}/(\sqrt{z}+\sqrt{z-1}))^{-(\al+\la)}$ is the Szeg\"{o} function, such that $D_{+}(x)D_{-}(x)=x^{\al+\la}$ for $x\in (0,1).$ The branches in the above are chosen as $\arg z\in (-\pi, \pi)$ and $\arg(z-1) \in (-\pi, \pi).$
\par
Then
\bq\label{R15}
S(z){P^{(\infty)}}^{-1}(z)=I+\mathcal{O}\left(n^{-1}\right),\;\;n\rw\infty,
\eq
where the error term is uniform for $z$ away from the end points $0,$ $1$ and $t\in(0,c],$ $c$ is a positive and finite constant.
Due to the factor $a(z)$ in $(\ref{R14})$ with fourth root singularities at $z=0,1,$  the jump matrices $S(z){P^{(\infty)}}^{-1}(z)$ are not uniformly close to the unit matrix in the neighborhood of $0$ and $1.$ So, it's need to construct the local parametrices in the neighborhoods of these points.

\begin{rem}
Let
\bq\label{K1}
N(z)=P^{(\infty)}(z)\left(\frac{z+\frac{t}{4n}}{z}\right)^{-\frac{\la}{2}\sigma_{3}},
\eq
where $\arg z \in (-\pi, \pi),$ $\arg(z+t/4n) \in (-\pi,\pi),$ and $P^{(\infty)}(z)$ is given by (\ref{R14}). Then $N(z)$ satisfies the same RH problem for $P^{(\infty)}(z)$ and the jump condition $(\ref{RE16})$ replaced by
\bq\label{K2}
N_{+}(x)=N_{-}(x)\left\{
\begin{array}{lll}
e^{i\la\pi\sigma_{3}},& \; x\in (-\frac{t}{4n}, 0),\\
\\
\left(
\begin{matrix}
0&x^{\alpha}\left(x+\frac{t}{4n}\right)^{\lambda}\\
-x^{-\alpha}\left(x+\frac{t}{4n}\right)^{-\lambda}&0\\
\end{matrix}
\right),& \; x\in (0,1).\\
\end{array}
\right.
\eq
\end{rem}
Moreover, from (\ref{R14}) and (\ref{K1}), one finds,
\bq\label{K3}
P^{(\infty)-1}(z)N(z)=I+\mathcal{O}\left(n^{-1}\right),\;\;z\in \mathbb{C}\setminus \{U(0,r)\cup U(1,r)\},\;\;{\rm and}\;\; r>t/4n,
\eq
where the error term is uniform for $z\in \mathbb{C}\setminus \{U(0,r)\cup U(1,r)\},$  $r>t/4n,$ $t\in (0,c]$ and $c$ is a positive constant.
\par
\subsection{Local parametrix $P^{(0)}(z)$ at $z=0$}
\par
We construct the local parametrix $P^{(0)}(z)$ in the neighborhood $U(0,r)=\{z\in \mathbb{C}: |z|<r\}$ for small $r,$ and $r>t/4n.$ The parametrix $P^{(0)}(z)$ satisfies a RH problem as follows
\par
$(a)$ $P^{(0)}(z)$ is analytic in $U(0,r)\setminus \{\bigcup_{k=1}^{3}\Sigma_{k}\},$ see Figure 2.
\par
$(b)$ $P^{(0)}(z)$ has the same jump conditions with $S(z)$ on $U(0,r)\bigcap \Sigma_{k},k=1,2,3,$ see (\ref{RE14}).
\par
$(c)$ For $z \in \partial U(0,r),$ $P^{(0)}(z)$ satisfies the following matching condition,
\par
\bq\label{R16}
P^{(0)}(z){P^{(\infty)}}^{-1}(z)=I+\mathcal{O}\left(n^{-\frac{1}{2}}\right), \;\;n \rw \infty.
\eq
\par
$(d)$ The asymptotic behavior of $P^{(0)}(z)$ at $z=0$ is the same as $S(z)$ in (\ref{RE15}).
\par
In order to convert the jump of the $P^{(0)}(z)$ to constant jump, a transformation is defined as
\bq\label{RW0}
\widetilde{P}^{(0)}(z)=P^{(0)}(z)e^{-n\phi(z)\sigma_{3}}(-z)^{\frac{\alpha+\lambda}{2}\sigma_{3}}\left(\frac{-\left(z+\frac{t}{4n}\right)}{-z}\right)^{\frac{\lambda}{2}\sigma_{3}}, \;\;z \in U(0,r)\setminus \Sigma_{k},\;k=1,2,3,
\eq
where $\arg(-z)\in (-\pi, \pi),$ $\arg\left(-\left(z+t/4n\right)\right)\in (-\pi,\pi),$ and $\phi(z)$ is given by (\ref{RW01}). Here, we set that $\arg(-z)=\mp\pi$ on the positive and negative side of $\Sigma_{2},$ respectively, and $\arg\left(-\left(z+t/4n\right)\right)=\mp\pi$ as $z$ is form above and below of $\left(-t/4n,0\right),$ respectively.
\par
Obviously, $\widetilde{P}^{(0)}(z)$ satisfies the following RH problem.
\par
$(a)$ $\widetilde{P}^{(0)}(z)$ is analytic for $z\in U(0,r)\setminus \{\cup_{k=1}^{3}\gamma_{k}\}\cup \left(-\frac{t}{4n},0\right),$ see Figure 3.
\par
$(b)$  $\widetilde{P}^{(0)}(z)$ fulfills the following constant jumps,
\bq\label{RE17}
\widetilde{P}_{+}^{(0)}(z)=\widetilde{P}_{-}^{(0)}(z)\left\{
\begin{array}{llll}
e^{-{i}\lambda\pi{\sigma_{3}}}, & z\in \left(-\frac{t}{4n},0\right),\\
\\
\left(
\begin{matrix}
1&0\\
e^{-i(\lambda+\alpha)\pi}&1\\
\end{matrix}
\right),& z \in\gamma_{3}\bigcap U(0,r),\\
\\
\left(
\begin{matrix}
0&1\\
-1&0\\
\end{matrix}
\right),& z \in \gamma_{2}\bigcap U(0,r),\\
\\
\left(
\begin{matrix}
1&0\\
e^{i(\lambda+\alpha)\pi}&1\\
\end{matrix}
\right),& z\in \gamma_{1}\bigcap U(0,r).
\end{array}
\right.
\eq
\par
$(c)$ The asymptotic behavior of $\widetilde{P}^{(0)}(z)$ at the origin in the following four domains $\Omega_{k}, k=1,\ldots,4,$ illustrated in Figure 3,
\\
$(c_{1})$ $z\in \Omega_{1},$
\bbq
\widetilde{P}^{(0)}(z)=\mathcal{O}(1)e^{-n\phi(z)\sigma_{3}}(-z)^{\frac{\alpha}{2}\sigma_{3}}\left(-\left(z+t/4n\right)\right)^{\frac{\lambda}{2}\sigma_{3}},\;\;z\rw 0,
\eeq
\\
$(c_{2})$ $z\in \Omega_{2},$
\bbq
\widetilde{P}^{(0)}(z)=\mathcal{O}(1)e^{-n\phi(z)\sigma_{3}}(-z)^{\frac{\alpha}{2}\sigma_{3}}\left(-\left(z+t/4n\right)\right)^{\frac{\lambda}{2}\sigma_{3}}\left(
\begin{matrix}
1&0\\
e^{i(\lambda+\alpha)\pi}&1\\
\end{matrix}
\right),\;\;z\rw 0,
\eeq
\\
$(c_{3})$ $z\in \Omega_{3},$
\bbq
\widetilde{P}^{(0)}(z)=\mathcal{O}(1)e^{-n\phi(z)\sigma_{3}}(-z)^{\frac{\alpha}{2}\sigma_{3}}\left(-\left(z+t/4n\right)\right)^{\frac{\lambda}{2}\sigma_{3}}\left(
\begin{matrix}
1&0\\
-e^{-i(\alpha+\lambda)\pi}&1\\
\end{matrix}
\right),\;\;z\rw 0,
\eeq
\\
$(c_{4})$ $z\in \Omega_{4},$
\bbq
\widetilde{P}^{(0)}(z)=\mathcal{O}(1)e^{-n\phi(z)\sigma_{3}}(-z)^{\frac{\alpha}{2}\sigma_{3}}\left(-\left(z+t/4n\right)\right)^{\frac{\lambda}{2}\sigma_{3}},\;\;z\rw 0.
\eeq
\par
($d$) The behavior of $\widetilde{P}^{(0)}(z)$ at $z=-\frac{t}{4n}$ is
\bbq
\widetilde{P}^{(0)}(z)=\mathcal{O}(1)\left(-\left(z+t/4n\right)\right)^{\frac{\lambda}{2}\sigma_{3}}.
\eeq

\begin{center}
\begin{tikzpicture}
\begin{scope}[line width=2pt]
\draw[->,>=stealth] (0,0)--(3,0);
\draw[-] (0,0)--(6,0);
\draw[->,>=stealth] (-2.5,0)--(-1,0);
\draw[-] (-2.5,0)--(0,0);
\draw[dashed] (-5,0)--(-2.5,0);
\draw[-] (0,0)--(2.4,-2.88);
\draw[-] (0,0)--(2.4,2.88);
\draw[->,>=stealth] (0,0)--(1.485,1.8);
\draw[->,>=stealth] (0,0)--(1.485,-1.8);
\node[below] at (-0.05,0) { $O$};
\node[below] at (-2.5,0) {{$-\frac{t}{4n}$}};
\node[above] at (1.2,1.6) {{$\gamma_{3}$}};
\node[above] at (2,0) {{$\gamma_{2}$}};
\node[above] at (1.7,-1.8) {{$\gamma_{1}$}};

\node[above] at (-1,.6) {{$\Omega_{4}$}};
\node[above] at (3,1) {{$\Omega_{3}$}};
\node[above] at (3.2,-1.5) {{$\Omega_{2}$}};
\node[above] at (-1,-1.5) {{$\Omega_{1}$}};
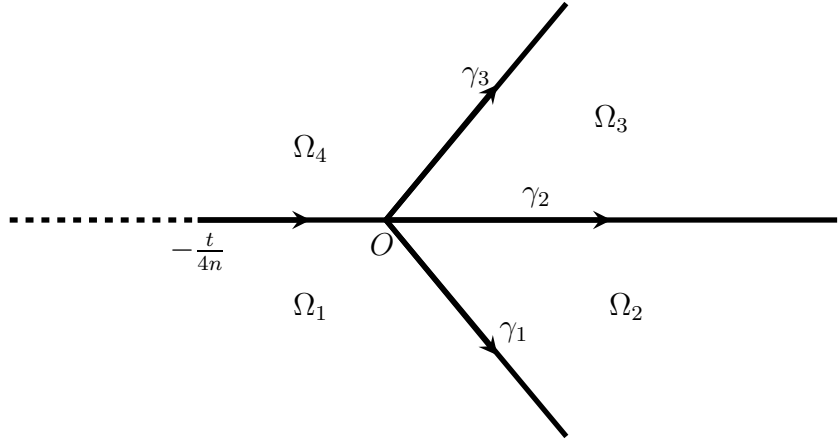
\node[below] at (-.2,-3) { Figure 3. Counters for the RH problem of $\widetilde{P}^{(0)}(z)$ in the z plane};
\end{scope}
\end{tikzpicture}
\end{center}

\par
From (\ref{RW01}), $\xi$ is defined as
\bq\label{RW02}
\xi=n^{2}\phi^{2}(z)=-16n^{2}z\left(1-\frac{z}{3}-\frac{z^{2}}{45}+\mathcal{O}\left(z^{3}\right)\right),
\eq
is a conformal mapping for $z\in U(0,r),$ and $r$ is small.
In fact, with the jump condition (\ref{RE17}) and the matching condition (\ref{R16}), $\widetilde{P}^{(0)}(z)$ satisfies the following equation,
\bq\label{RW03}
\widetilde{P}^{(0)}(z)=E(z)\Phi(n^{2}\phi^{2}(z), 4nt)e^{-\frac{i\pi}{2}\sigma_{3}},
\eq
where the factor $e^{-\frac{i\pi}{2}\sigma_{3}}$ compensates for the reversed orientation of $\xi\sim-16n^{2}z,$ see (\ref{RW02}) and (\ref{RE21}), $E(z)$ is given by
\bq\label{T2}
E(z)=N(z)e^{\frac{i\pi}{2}\sigma_{3}}\left(-z\right)^{\frac{\al+\la}{2}\sigma_{3}}\left(\frac{-\left(z+\frac{t}{4n}\right)}{-z}\right)^{\frac{\la}{2}\sigma_{3}}\frac{I-i\sigma_{1}}{\sqrt{2}}\left(n^{2}\phi^{2}(z)\right)^{\frac{1}{4}\sigma_{3}},
\eq
where $N(z)$ is given by (\ref{K1}), $\arg(-z)\in (-\pi, \pi)$ and $\arg(n^{2}\phi^{2}(z))\in (-\pi, \pi).$  From (\ref{K1}), (\ref{R16})-(\ref{RW0}), (\ref{RW03})-(\ref{T2}) and (\ref{RE21}), one can vertify the matching condition (\ref{R16}), see also Remark \ref{R00}.
\par
$E(z)$ is analytic in $U(0,r).$ Since $E(z)$ is analytic in $U(0,r)\setminus (-t/4n, 0)\cup\gamma_{2}.$ It derives directly that $E_{+}(x)=E_{-}(x)$ for $x\in (-t/4n, 0)\cup\gamma_{2}$ from (\ref{R14}), (\ref{K1}) and (\ref{T2}), and $E(z)$ has a weak singularity at origin .
\par
\subsection{Local parametrix $P^{(1)}(z)$ at $z=1$}
For $z\in U(1,r),$ $U(1,r)=\{z:|z-1|<r\},$ and $r$ is sufficiently small and positive, then the local parametric $P^{(1)}(z)$ satisfies the following RH problem.
\par
$(a)$ $P^{(1)}(z)$ is analytic in $U(1,r)\setminus \Sigma_{S},$ $\Sigma_{S}$ denotes the contour of $S,$ see Figure 2.
\par
$(b)$ $P^{(1)}(z)$ shares the jump conditions of $S(z)$ in $U(1,r),$ see (\ref{RE14}).
\par
$(c)$ For $z \in \partial U(1,r),$ $P^{(1)}(z)$ satisfies the matching condition as follows
\bq\label{R17}
P^{(1)}(z)P^{(\infty)-1}(z)=I+\mathcal{O}\left(n^{-1}\right),\; {\rm as}\;\; n\rw \infty.
\eq
In fact, $P^{(1)}(z)$ can be expressed in terms of the Airy function and its derivatives, the detail presented in Sect.5 below.
\subsection{Riemann-Hilbert problem for $R$}
\par
Here, we introduce the last transformation $S\rw R,$ and $R(z)$ is defined as follows
\bq\label{RW05}
R(z)=\left\{
\begin{array}{llll}
S(z)P^{(\infty)-1}(z),& z \in \mathbb{C}\setminus\{U(0,r)\cup U(1,r)\cup \Sigma_{S}\} ,\\
\\
S(z)P^{(0)-1}(z),& z\in U(0, r)\setminus \Sigma_{S},\\
\\
S(z)P^{(1)-1}(z),& z\in U(1, r)\setminus \Sigma_{S}.\\
\end{array}
\right.
\eq
Hence, $R(z)$ satisfies the following RH problem.
\par
$(a)$ $R(z)$ is analytic in $\mathbb{C}\setminus \Sigma_{R}$, see Figure 4.
\par
$(b)$ $R(z)$ satisfies the jump conditions
\bq\label{Rw06}
R_{+}(z)=R_{-}(z)\left\{
\begin{array}{llll}
P^{(\infty)}(z)J_{S}(z)P^{(\infty)-1}(z),& z \in \Sigma_{R}\setminus \{\partial U(0,r)\cup \partial U(1,r)\}, \\
\\
P^{(0)}(z)P^{(\infty)-1}(z),& z\in \partial U(0, r),\\
\\
P^{(1)}(z)P^{(\infty)-1}(z),& z\in \partial U(1, r),\\
\end{array}
\right.
\eq
where $J_{S}$ denotes the jump matrices in (\ref{RE14}).
\par
$(c)$ For $z\rw \infty,$ $R(z)=I+\mathcal{O}\left(z^{-1}\right).$
\par
By (\ref{RE14}), (\ref{Rw06}) and the matching conditions on the boundaries (\ref{R15}), (\ref{R16}), (\ref{R17}) and $\phi(z)$ in (\ref{RW01}), one finds,
\bq\label{Rw07}
J_{R}(z)=\left\{
\begin{array}{llll}
I+\mathcal{O}\left(n^{-\frac{1}{2}}\right),& z\in \partial U(0, r)\cup \partial U(1,r),\\
\\
I+\mathcal{O}\left(n^{-1}\right),& z \in \widetilde{\Sigma}_{2},\\
\\
I+\mathcal{O}\left(e^{-cn}\right),& z \in \Sigma_{R}\setminus \{\partial U(0,r)\cup \partial U(1,r)\cup \widetilde{\Sigma}_{2}\},\\
\end{array}
\right.
\eq
where $c$ is a positive constant, and the error term is uniform for $z \in \Sigma_{R}.$ Then one finds,
\bq\label{Rw08}
\|J_{R}(z)-I\|_{L^{2}\cap L^{\infty}(\Sigma_{R})}=\mathcal{O}\left(n^{-\frac{1}{2}}\right).
\eq
Furthermore,
\bq\label{Rw09}
R(z)=I+\frac{1}{i2\pi}\int_{\Sigma_{R}}\frac{R_{-}(\tau)\left(J_{R}(\tau)-I\right)}{\tau-z}d\tau,\;z\notin \Sigma_{R},
\eq
by the method and procedure of norm estimation of Cauchy operator as show in \cite{DeiftBook, DX1999}, one has,
\bq\label{Rw10}
R(z)=I+\mathcal{O}\left(n^{-\frac{1}{2}}\right),
\eq
where the error term is uniform for $z\in \mathbb{C}.$
\begin{rem}\label{R00}
In the double scaling process, let $n\rw \infty,$ $t\rw 0,$ and $s=4nt$ such that $s\in (0,\infty).$ In fact $t$ can be taken in the interval $(0,c]$ with $c$ is finite and positive, for $s\rw \infty,$ then
the error term (\ref{Rw07}) uniformly $\mathcal{O}\left(n^{-\frac{1}{2}}\right)$ for $z\in \mathbb{C}.$ This can be verified from calculations, see (\ref{RE113}).
\end{rem}
\par
It now completes the nonlinear steepest descent analysis of $Y\rw T\rw S\rw R.$ By a list of inverse transformations, we will show
the large-$n$ asymptotic behavior of the kernel in the below sections.

\begin{center}
\begin{tikzpicture}
\begin{scope}[line width=2pt]
\draw(0,0) circle(0.8);
\draw(7,0) circle(0.8);

\draw[->,>=stealth] (0.8,0)--(3,0);
\draw[-] (2.6,0)--(6.2,0);

\draw[->,>=stealth] (7.8,0)--(9,0);
\draw[->,>=stealth] (3.4,1.6)--(3.5,1.6);
\draw[->,>=stealth] (3.4,-1.6)--(3.5,-1.6);

\draw[-] (7.8,0)--(11,0);
\draw (0.567,0.567) .. controls (3.4,1.9) and (3.6,2) .. (6.433,.567);
\draw (0.567,-0.567) .. controls (3.4,-1.9) and (3.6,-2) .. (6.433,-.567);

\node[above] at (2.8,1.7) {$\widetilde{\Sigma}_{3}$};
\node[above] at (3.8,0) {$\widetilde{\Sigma}_{2}$};
\node[above] at (2.9,-1.6) {$\widetilde{\Sigma}_{1}$};
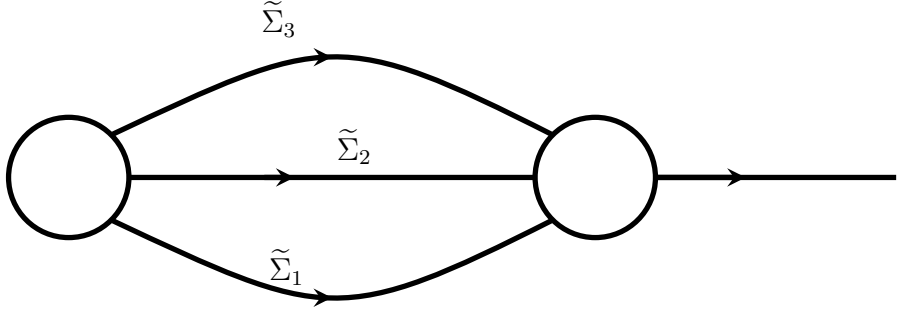
\node[below] at (4.6,-2) { Figure 4. The contour  for $R(z).$};
\end{scope}
\end{tikzpicture}
\end{center}

\par
\subsection{Proof of Theorem 1}
In this subsection, we prove that the Painlev\'e \uppercase\expandafter{\romannumeral5} Kernel at the hard edge.
\begin{proof}
\par
The kernel in (\ref{00a1}) associated with the perturbed Laguerre weight $w(x,t)$ in (\ref{R1}) denoted as $K_{n}(x,y;t)$ and by the Christoffel--Darboux formula \cite{Szego1939},  it has a simple closed form as follows
\begin{equation}\label{T4}
K_{n}(x,y; t)=\left(w(x,t)\right)^{\frac{1}{2}}\left(w(y,t)\right)^{\frac{1}{2}}\frac{\pi_{n}(x)\pi_{n-1}(y)-\pi_{n}(y)\pi_{n-1}(x)}{(x-y)h_{n-1}},
\end{equation}
where $h_{n-1}$ is the square of the $L^{2}$ norm, see (\ref{T6}). The above kernel can also expressed by $Y(z)$ in (\ref{RE4}),
\bq\label{T7}
K_{n}(x,y; t)=\frac{\left(w(x,t)w(y,t)\right)^{\frac{1}{2}}}{i2\pi(x-y)}\left(Y_{+}^{-1}(y)Y_{+}(x)\right)_{21}.
\eq
From the first transform (\ref{RW2}) and (\ref{R24}), the support of the equilibrium measure is $[0, 4n]$ and $x, y \in [0,4n]$ in the kernel (\ref{T7}). In order to normalize the the support of the equilibrium measure $[0, 4n]$ as $[0,1],$ so we consider the following re-scaled kernel,
\begin{align}\label{RE138}
4nK_{n}(4nx,4ny; t)&=\frac{\left(w(4nx, t)w(4ny, t)\right)^{\frac{1}{2}}}{i2\pi(x-y)}\left(Y_{+}^{-1}(4ny)Y_{+}(4nx)\right)_{21}\nonumber\\
&=\frac{1}{i2\pi(x-y)}\left(
\begin{matrix}
0&1\\
\end{matrix}
\right)\left(w(4ny, t)\right)^{\frac{1}{2}}Y_{+}^{-1}(4ny)Y_{+}(4nx)w(4nx, t)^{\frac{1}{2}}
\left(
\begin{matrix}
1\\
0\\
\end{matrix}
\right).
\end{align}
\par
Taking the inverse transformations from $Y$ to $R,$ with the aid of (\ref{RW2}), (\ref{RE13}), (\ref{RW0}) and (\ref{RW03}), one finds,
\bq\label{Rw11}
Y_{+}(4nx)w^{\frac{1}{2}}(4nx, t)\left(
\begin{matrix}
1\\
0\\
\end{matrix}
\right)
=M_{0}R(x)E(x)\Phi_{-}(n^{2}\phi^{2}(x),s)e^{\frac{i\pi(\al+\la-1)}{2}\sigma_{3}}
\left(
\begin{matrix}
1\\
1\\
\end{matrix}
\right),
\eq
and
\bq\label{Rw12}
\left(
\begin{matrix}
0&1\\
\end{matrix}
\right)w^{\frac{1}{2}}(4ny)Y^{-1}_{+}(4ny, t)=\left(
\begin{matrix}
-1&1\\
\end{matrix}
\right)
e^{-\frac{i\pi(\al+\la-1)}{2}\sigma_{3}}\Phi^{-1}_{-}(n^{2}\phi^{2}(y),s)E^{-1}(y)R^{-1}(y)M^{-1}_{0},
\eq
where $M_{0}=(-1)^{n}(4n)^{\left(n+\frac{\al+\la}{2}\right)\sigma_{3}}e^{\frac{n\ell}{2}\sigma_{3}},$ $s=4nt$ is finite, and one has been used the fact
\bbq
e^{n\left(g_{+}(x)+\phi_{+}(x)-\frac{\ell}{2}\right)}=e^{2nx+in\pi}=(-1)^{n}e^{2nx},
\eeq
which follows from (\ref{a4}) and (\ref{a5}).
\par
Let
\bq\label{Rw13}
\left(
\begin{matrix}
\varphi_{1}\left(n^{2}\phi^{2}(x),s\right)\\
\varphi_{2}\left(n^{2}\phi^{2}(x),s\right)\\
\end{matrix}
\right)
=\Phi_{-}(n^{2}\phi^{2}(x),s)e^{\frac{i\pi(\al+\la-1)}{2}\sigma_{3}}
\left(
\begin{matrix}
1\\
1\\
\end{matrix}
\right),
\eq
by (\ref{RE138}), (\ref{Rw11}) and (\ref{Rw12}), then the kernel can be rewritten as
\begin{align}\label{RR01}
4nK_{n}(4nx,4ny; s)
&=\frac{\left(
\begin{matrix}
-\varphi_{2}\left(n^{2}\phi^{2}(y),s\right)\\
\varphi_{1}\left(n^{2}\phi^{2}(y),s\right)\\
\end{matrix}
\right)^{T}E^{-1}(y)R^{-1}(y)R(x)E(x)\left(
\begin{matrix}
\varphi_{1}\left(n^{2}\phi^{2}(x),s\right)\\
\varphi_{2}\left(n^{2}\phi^{2}(x),s\right)\\
\end{matrix}
\right)}{i2\pi(x-y)},
\end{align}
where $Z^{T}$ represents the transpose of $Z$. By (\ref{RW02}), the variable $x,$ and $y$ rescaled as $x=\frac{u}{16n^{2}},$ $y=\frac{{\rm v}}{16n^{2}},$ and $u, {\rm v} \in (0,+\infty),$ then we have,
\bq\label{Rw14}
n^{2}\phi^{2}(x)=-u\left(1+\mathcal{O}\left(n^{-2}\right)\right),\;\;n^{2}\phi^{2}(y)=-{\rm v}\left(1+\mathcal{O}\left(n^{-2}\right)\right).
\eq
While $\varphi_{k}\left(\xi,s\right),$ $\xi\in(-\infty,0),$ $k=1,2,$  which can be extended as analytic functions, so one finds,
\bq\label{Rw15}
\varphi_{k}\left(n^{2}\phi^{2}(x),s\right)=\varphi_{k}\left(-u,s\right)+\mathcal{O}\left(n^{-2}\right),\;\;\varphi_{k}\left(n^{2}\phi^{2}(y),s\right)=\varphi_{k}\left(-{\rm v},s\right)+\mathcal{O}\left(n^{-2}\right),\;k=1,2.
\eq
Further more, $E(z)$ is analytic in $U(0,r),$ one finds,
\bq\label{Rw16}
E^{-1}(y)E(x)=I+\mathcal{O}\left(x-y\right)=I+\mathcal{O}\left(n^{-2}\right),
\eq
and $R(z)$ is also analytic in $U(0,r),$ and
\bq\label{Rw17}
R^{-1}(y)R(x)=I+\mathcal{O}\left(x-y\right)=I+\mathcal{O}\left(n^{-2}\right).
\eq
Substituting (\ref{Rw14}), (\ref{Rw15}), (\ref{Rw16}) and (\ref{Rw17}) into (\ref{RR01}), then the re-scaled kernel for large $n$ as follows
\bq\label{Rw18}
\frac{1}{4n}K_{n}(\frac{u}{4n},\frac{{\rm v}}{4n}; s)=\frac{\varphi_{1}(-{\rm v},s)\varphi_{2}(-u,s)-\varphi_{1}(-u,s)\varphi_{2}(-{\rm v},s)}{i2\pi(u-{\rm v})}+\mathcal{O}\left(\frac{1}{n^{2}}\right),
\eq
where the error term is uniform for $u,{\rm v}\in (0,+\infty)$ and $s\in (0,+\infty).$ As $n\rw \infty,$ then (\ref{T1}) follows from (\ref{Rw18}).
\par
By the unique solution of the RH problem (\ref{RE20}), (\ref{RE21}), (\ref{RE22}) and (\ref{RE23}) for $\Phi(\xi,s)$ in the $\xi$ plane, we extend (\ref{Rw13}) as follows
\bq\label{T3}
\left(
\begin{matrix}
\varphi_{1}\left(\xi,s\right)\\
\varphi_{2}\left(\xi,s\right)\\
\end{matrix}
\right)
=\Phi(\xi,s)e^{\frac{i\pi(\al+\la-1)}{2}\sigma_{3}}
\left(
\begin{matrix}
1\\
1\\
\end{matrix}
\right),
\eq
where $\xi\in \widehat{\Omega}_{3},$ see Figure 1.
Substituting (\ref{T3}) into the Lax pair (\ref{RE24}) and (\ref{RE25}) for $\Phi(\xi, s),$ and eliminate $\varphi_{1}(\xi,s)$ or $\varphi_{2}(\xi,s),$ then it follows the second order differential equation in (\ref{R11}). By the initial conditions of $r(0)$ and $r'(0)$ are in (\ref{R7}), (\ref{T26}), respectively, and set $s=0$ in (\ref{R11}), then the differential equation reduces to (\ref{R13}).
\end{proof}

\section{Painlev\'e \uppercase\expandafter{\romannumeral5} kernel to Bessel kernels as $s\rightarrow 0^{+}$ and $s \rw \infty$}
In this section, we show that the Painlev\'e \uppercase\expandafter{\romannumeral5} kernel (\ref{T1}) translates to the Bessel kernel $\mathbf{J}_{\al+\la}$ and the Bessel kernel $\mathbf{J}_{\al}$ as $s\rightarrow 0^{+}$ and $s\rightarrow \infty,$ respectively.
\par
\subsection{Painlev\'e \uppercase\expandafter{\romannumeral5} kernel to the Bessel kernel $\mathbf{J}_{\al+\la}$ as $s\rightarrow 0^{+}$}
\par
If $s\rw 0,$ the factor $(\xi-s)^{\la}$ merges $\xi^{\al}$ as $\xi^{\al+\la}$ in the RH problem for $\Phi(\xi,s),$ see $(\ref{RE20})$-$(\ref{RE23}),$ and the jump on the interval $(0,s)$ vanishes. Hence, the RH problem for $\Phi(\xi, s)$ can be approximated by the following limiting RH problem for $\Phi_{0}(\xi,s).$
\par
$(a)$ $\Phi_{0}(\xi)$ is analytic in $\mathbb{C}\setminus \displaystyle\cup_{j=1}^{3}\widehat{\Sigma}_{j}',$ and the contour see Figure 5.
\par
$(b)$ $\Phi_{0}(\xi)$ fulfills the jump relation
\bq\label{RE57}
(\Phi_{0})_{+}(\xi)=(\Phi_{0})_{-}(\xi)\left\{
\begin{array}{llll}
\left(
\begin{matrix}
1&0\\
e^{\pi(\lambda+\alpha)i}&1\\
\end{matrix}
\right),& \xi\in \widehat{\Sigma}_{1}',\\
\\
\left(
\begin{matrix}
0&1\\
-1&0\\
\end{matrix}
\right),& \xi \in \widehat{\Sigma}_{2}',\\
\\
\left(
\begin{matrix}
1&0\\
e^{-\pi(\lambda+\alpha)i}&1\\
\end{matrix}
\right),& \xi \in\widehat{\Sigma}_{3}'.
\end{array}
\right.
\eq
\par
$(c)$ For $\xi\rightarrow \infty,$
\bq\label{RE58}
\Phi_{0}(\xi)=\left(I+\mathcal{O}\left(\frac{1}{\xi}\right)\right)\xi^{-\frac{1}{4}\sigma_{3}}\frac{I+i\sigma_{1}}{\sqrt{2}}e^{\sqrt{\xi}\sigma_{3}},
\eq
where the argument takes as $\arg{\xi} \in (-\pi, \pi).$

\begin{center}
\begin{tikzpicture}
\begin{scope}[line width=2pt]
\draw(0,0) circle(0.9);
\draw[->,>=stealth] (-4,0)--(-3,0);
\draw[-] (-5.8,0)--(-0.9,0);

\draw[dashed] (0.9,0)--(4.6,0);

\draw[->,>=stealth] (-1.8,3.116)--(-1.35,2.337);
\draw[-] (-2.7,4.674)--(-0.45,0.779);
\draw[->,>=stealth] (-1.8,-3.116)--(-1.35,-2.337);
\draw[-] (-2.7,-4.674)--(-0.45,-0.779);
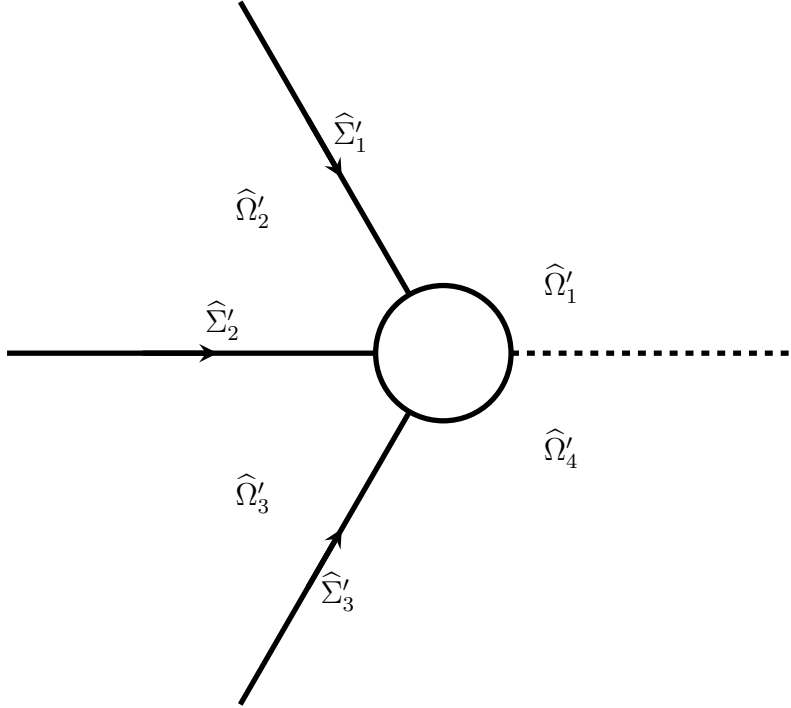
\node[below] at (-.8,-5) {Figure 5. Contours and regions};

\node[above] at (-1.4,-3.6) {{$\widehat{\Sigma}_{3}'$}};
\node[above] at (-3,0) {{ $\widehat{\Sigma}_{2}'$}};
\node[above] at (-1.3,2.5) {{ $\widehat{\Sigma}_{1}'$}};

\node[above] at (1.5,.5) {{ $\widehat{\Omega}_{1}'$}};
\node[above] at (-2.6,1.5) {{ $\widehat{\Omega}_{2}'$}};
\node[above] at (-2.6,-2.3) {{ $\widehat{\Omega}_{3}'$}};
\node[above] at (1.5,-1.7) {{ $\widehat{\Omega}_{4}'$}};
\end{scope}
\end{tikzpicture}
\end{center}

\par
The modified Bessel functions are used to construct the local parametrix at origin, we refer \cite{KM2004}, more information about the modified Bessel function can be found in \cite{B1953, OLC2010}. The unique solution of the RH problem (\ref{RE57})-(\ref{RE58}) for $\Phi_{0}(\xi)$ can be constructed in terms of the modified Bessel functions $I_{\upsilon}(z)$ and $K_{\upsilon}(z)$ as follows
\bq\label{RE59}
\Phi_{0}(\xi)=M_{0}\left\{
\begin{array}{llll}
\left(
\begin{matrix}
I_{\alpha+\lambda}\left(\xi^{\frac{1}{2}}\right)&i\frac{1}{\pi}K_{\alpha+\lambda}\left(\xi^{\frac{1}{2}}\right)\\
i\pi\xi^{\frac{1}{2}}I'_{\alpha+\lambda}\left(\xi^{\frac{1}{2}}\right)&-\xi^{\frac{1}{2}}K'_{\alpha+\lambda}\left(\xi^{\frac{1}{2}}\right)
\end{matrix}
\right), &\xi \in \widehat{\Omega}_{1}'\cup\widehat{\Omega}_{4}',\\
\\
\left(
\begin{matrix}
I_{\alpha+\lambda}\left(\xi^{\frac{1}{2}}\right)&i\frac{1}{\pi}K_{\alpha+\lambda}\left(\xi^{\frac{1}{2}}\right)\\
i\pi\xi^{\frac{1}{2}}I'_{\alpha+\lambda}\left(\xi^{\frac{1}{2}}\right)&-\xi^{\frac{1}{2}}K'_{\alpha+\lambda}\left(\xi^{\frac{1}{2}}\right)
\end{matrix}
\right)\left(
\begin{matrix}
1&0\\
-e^{i(\lambda+\alpha)\pi}&1\\
\end{matrix}
\right),& \xi\in \widehat{\Omega}_{2}',\\
\\
\left(
\begin{matrix}
I_{\alpha+\lambda}\left(\xi^{\frac{1}{2}}\right)&i\frac{1}{\pi}K_{\alpha+\lambda}\left(\xi^{\frac{1}{2}}\right)\\
i\pi\xi^{\frac{1}{2}}I'_{\alpha+\lambda}\left(\xi^{\frac{1}{2}}\right)&-\xi^{\frac{1}{2}}K'_{\alpha+\lambda}\left(\xi^{\frac{1}{2}}\right)
\end{matrix}
\right)\left(
\begin{matrix}
1&0\\
e^{-i(\lambda+\alpha)\pi}&1\\
\end{matrix}
\right),& \xi \in \widehat{\Omega}_{3}',
\end{array}
\right.
\eq
where the constant matrix $M_{0}$ is given by
\bq\label{RE60}
M_{0}=
\left(\left(4(\alpha+\lambda)^{2}+3\right)i\sigma_{-}/8+I\right)\pi^{\frac{1}{2}\sigma_{3}}.
\eq
From (\ref{RE59}), $\Phi_{0}(\xi)$ has the following asymptotic behavior at infinity,
\bq\label{A1}
\Phi_{0}(\xi)
=\left(I+C_{0}\xi^{-1}
+\mathcal{O}\left(\xi^{-\frac{3}{2}}\right)\right)\xi^{-\frac{1}{4}\sigma_{3}}\frac{I+i\sigma_{1}}{\sqrt{2}}e^{\sqrt{\xi}\sigma_{3}},\;\; \xi \rw \infty,
\eq
where
\bq\label{A2}
C_{0}=\frac{4(\alpha+\lambda)^{2}-1}{128}
\left(
\begin{matrix}
4(\alpha+\lambda)^{2}-9&i{16}\\
i\frac{(4(\alpha+\lambda)^2-9)(4(\alpha+\lambda)^2-13)}{12}&9-4(\alpha+\lambda)^{2}
\end{matrix}
\right).
\eq
These can be verified by the following asymptotic behavior of the modified Bessel functions \cite{AAR1999}, see also \cite{B1953, OLC2010},
\bbq\label{RE71}
K_{\up}(z)\sim \sqrt{\frac{\pi}{2z}}e^{-z}\left(1+\sum_{j=1}^{\infty}\frac{(-1)^{j}(\up+1/2)_{j}(-\up+1/2)_{j}}{(2z)^{j}j!}\right),\;\;\left|\arg{z}\right|<\frac{3\pi}{2},
\eeq
\bbq\label{RE72}
I_{\up}(z)\sim \frac{e^{z}}{\sqrt{2\pi{z}}}\sum_{j=0}^{\infty}\frac{(\up+1/2)_{j}(-\up+1/2)_{j}}{(2z)^{j}j!},\;\; -\frac{\pi}{2}<\arg{z}<\frac{3\pi}{2},
\eeq
where
\bbq\label{RE73}
(a)_{n}=a(a+1)(a+2)\ldots (a+n+1),\;\; {\rm and}\;\;(a)_{0}=1.
\eeq
\begin{rem}
Comparing (\ref{RE21}) with (\ref{A1}), we note that $C_{1}(s)\left.\right|_{s=0}=C_{0}$ provides the initial values of functions $r(s),$ $q(s),$ and $t(s)$ in the Lax pair (\ref{RE24})-(\ref{RE25}), then one finds,
\bq\label{RE74}
r(0)=\frac{1-4(\alpha+\lambda)^{2}}{8},\;\; \; q(0)=\frac{(4(\alpha+\lambda)^{2}-1)(4(\alpha+\lambda)^{2}-9)}{128},
\eq

\bq\label{RE75}
t(0)=\frac{(4(\alpha+\lambda)^{2}-1)(4(\alpha+\lambda)^{2}-9)(4(\alpha+\lambda)^{2}-13)}{1536}.
\eq
If $\al+\la>0$ and $\al+\la\notin \mathbb{N},$ substituting the initial data (\ref{RE74}) into (\ref{RE30}) and let $s=0,$ then one finds,
\bq\label{R3}
r'(0)=-\frac{\la}{2(\al+\la)}.
\eq

\end{rem}
\par
We construct the local parametrix $\Phi_{1}(\xi,s)$ in $U(0,\va)$ with a fixed $\va>0,$ to approximate $\Phi(\xi,s)$  and match $\Phi_{0}(\xi)$ on the boundary $\partial{U(0,\va)}.$ By the different behaviors of the modified Bessel functions for $\al+\la \notin \mathbb{N}$ and $\al+\la \in \mathbb{N},$ we consider two cases separately.
\par
For $\al+\la \notin \mathbb{N},$ we split the modified Bessel matrix function in (\ref{RE59}) as follows
\bq\label{RE76}
\left(
\begin{matrix}
I_{\alpha+\lambda}\left(\xi^{\frac{1}{2}}\right)&i\frac{1}{\pi}K_{\alpha+\lambda}\left(\xi^{\frac{1}{2}}\right)\\
i\pi\xi^{\frac{1}{2}}I'_{\alpha+\lambda}\left(\xi^{\frac{1}{2}}\right)&-\xi^{\frac{1}{2}}K'_{\alpha+\lambda}\left(\xi^{\frac{1}{2}}\right)
\end{matrix}
\right)
=G(\xi)\xi^{\frac{\alpha+\lambda}{2}\sigma_{3}}\left(
\begin{matrix}
1&\frac{1}{i2\sin((\alpha+\lambda)\pi)}\\
0&1
\end{matrix}
\right),
\eq
where
\bq\label{RE77}
G(\xi)=\left(
\begin{matrix}
\xi^{-\frac{\alpha+\lambda}{2}}I_{\alpha+\lambda}\left(\xi^{\frac{1}{2}}\right)&\frac{i}{2\sin((\alpha+\lambda)\pi)}\xi^{\frac{\alpha+\lambda}{2}}I_{-(\alpha+\lambda)}\left(\xi^{\frac{1}{2}}\right)\\
i\pi\xi^{\frac{1-(\alpha+\lambda)}{2}}I'_{\alpha+\lambda}\left(\xi^{\frac{1}{2}}\right)&-\frac{\pi}{2\sin((\alpha+\lambda)\pi)}\xi^{\frac{1+\alpha+\lambda}{2}}I'_{-(\alpha+\lambda)}\left(\xi^{\frac{1}{2}}\right)
\end{matrix}
\right),
\eq
and $G(\xi)$ is an entire matrix function, one checks with the formulas,
\bbq\label{RE61}
K_{\upsilon}=\frac{\pi\left(I_{-\upsilon}(z)-I_{\upsilon}(z)\right)}{2\sin(\up\pi)},
\;\;
I_{\upsilon}\left(z\right)=\left(\frac{z}{2}\right)^{\upsilon}\sum_{j=0}^{\infty}\frac{1}{\Gamma(j+1)\Gamma(j+\upsilon+1)}\left(\frac{z^{2}}{4}\right)^{j},\;\; {\rm for}\;\;\up \notin \mathbb{N}.
\eeq
\par
With the aid of the decomposition in (\ref{RE76}), we construct $\Phi_{1}(\xi,s)$ in $U(0,\va)$ in the following and $s\ll \va$,
\bq\label{RE78}
\Phi_{1}(\xi,s)=G(\xi)\left(
\begin{matrix}
1&f(\xi)\\
0&1
\end{matrix}
\right)
\xi^{\frac{\al+\la}{2}\sigma_{3}}\left(1-\frac{s}{\xi}\right)^{\frac{\la}{2}\sigma_{3}}J,
\eq
where $\arg{\xi}\in (-\pi, \pi),$ $G(\xi)$ is an entire matrix function, see (\ref{RE77}), and the jump matrix $J$ is given by
\bq\label{RE79}
J=\left\{
\begin{array}{llll}
I, & \xi\in \widehat{\Omega}_{1}\cup\widehat{\Omega}_{4}, \\
\\
\left(
\begin{matrix}
1&0\\
-e^{i(\lambda+\alpha)\pi}&1\\
\end{matrix}
\right),& \xi \in \widehat{\Omega}_{2},\\
\\
\left(
\begin{matrix}
1&0\\
e^{-i(\lambda+\alpha)\pi}&1\\
\end{matrix}
\right),& \xi\in \widehat{\Omega}_{3},\\
\end{array}
\right.
\eq
and the sectors $\widehat{\Omega}_{j}, j=1,\ldots,4$ are described in Fig.1.
\par
Form the above expression of $\Phi_{1}(\xi,s)$ in (\ref{RE78}), then $\Phi_{1}(\xi,s)$ satisfies the following RH problem,
\par
$(a)$ $\Phi_{1}(\xi,s)$ is analysis in $U(0,\va)\setminus \displaystyle\cup_{j=1}^{3}\widehat{\Sigma}_{j}\cup\left(0,s\right),$ the contour see Figure 1.
\par
$(b)$ $\Phi_{1}(\xi,s)$ satisfies the same jump conditions in $U(0,\va)$ as $\Phi(\xi,s)$ in (\ref{RE20}).
\par
$(c)$ For $\xi \rightarrow 0,$ $\Phi_{1}(\xi,s)$ keeps the same asymptotic behavior of $\Phi(\xi,s)$ in (\ref{RE22}).
\par
$(d)$ For $\xi \rightarrow s,$ $\Phi_{1}(\xi,s)$ also has the same asymptotic behavior as $\Phi(\xi,s)$ in (\ref{RE23}).
\par
$(e)$ For $\xi\in \partial U(0,\va)$ and $s\rw 0,$ $\Phi_{1}(\xi,s)$ and $\Phi_{0}(\xi)$ satisfies the following matching condition,
\bq\label{RE80}
\Phi_{1}(\xi,s)\Phi_{0}(\xi)^{-1}=I+\mathcal{O}(s)+\mathcal{O}(s^{\al+\la+1}),\;\; \al+\la \notin \mathbb{N},\;\;\al+\la+1>0,
\eq
and
\bq\label{T11}
\Phi_{1}(\xi,s)\Phi_{0}(\xi)^{-1}=I+\mathcal{O}(s)+\mathcal{O}\left(s^{\al+\la+1}\log{s}\right),\;\;\al+\la \in \mathbb{N},\;\;\al+\la+1>0.
\eq
\par
Now we construct the scalar function $f(\xi)$ in (\ref{RE78}). The scalar function $f(\xi)$ is analytic in $U(0,\va)\setminus (-\va,0)$ and satisfies the following jump relation,
\bq\label{RE81}
f_{+}(\xi)-f_{-}(\xi)=\left|\xi\right|^{\al}\left|\xi-s\right|^{\la},\;\; \xi \in (-\va,0),
\eq
and it is convenient to define $f(\xi)$ as follows
\bq\label{RE82}
f(\xi)=-\frac{1}{4\pi\sin{(\alpha+\lambda)\pi}}\int_{\Gamma}\frac{z^{\alpha}(z-s)^{\lambda}}{z-\xi}dz,\;\; \xi \in U(0,\va)\setminus (-\va,0),\;\; \arg\xi\in (-\pi, \pi),
\eq
where the integration contour $\Gamma$ is defined as the upper and lower line segments of $[-\va_{2}, 0]$ and the circle $\left|z\right|=\va_{2}, \va_{2}>\va>0,$ see Figure 6.
\par
Moreover, to give an estimation
of $f(\xi)$ with the restrictions of $s<\epsilon<<\va_{1}=\left|\xi\right|\leq \va<\va_{2}.$ Furthermore, in order to use the Plemelj-Sokhotski formula, we split the original integration contour $\Gamma$ as three parts, see Figure 6, the first part is a closed integration path $\Gamma'$ which consists of the circle $\left|z\right|=\va_{2},$ the upper and lower line segments of $[-\va_{2},-\epsilon],$ and the circle $\left|z\right|=\epsilon,$ the second part is the circle $\left|z\right|=\epsilon,$ the third part is the upper and lower line segments of $[-\epsilon, 0].$ Then the explicit expression rewrites as follows
\begin{align}\label{RE83}
f(\xi)&=-\frac{1}{4\pi\sin{(\alpha+\lambda)\pi}}\int_{\Gamma'}\frac{z^{\alpha}(z-s)^{\lambda}}{z-\xi}dz
+\frac{1}{4\pi\sin{(\al+\la)\pi}}\int_{\left|z\right|=\epsilon}\frac{z^{\alpha}(z-s)^{\lambda}}{z-\xi}dz\nonumber\\
&\;\;\;\;-\frac{1}{4\pi\sin{(\alpha+\lambda)\pi}}\left(\int_{-\ep}^{0}\left(\frac{z^{\alpha}(z-s)^{\lambda}}{z-\xi}\right)_{+}dz+\int_{0}^{-\ep}\left(\frac{z^{\alpha}(z-s)^{\lambda}}{z-\xi}\right)_{-}dz\right)\nonumber\\
&=\frac{\xi^{\al}(\xi-s)^{\la}}{i2\sin{(\al+\la)\pi}}+\frac{i}{4\pi\sin{(\al+\la)\pi}}\lim_{\va'\rw0}\int_{-\pi+\va'}^{\pi-\va'}
\frac{(\ep{e^{i\theta}})^{\al+1}(\ep{e^{i\theta}}-s)^{\la}}{\ep{e^{i\theta}-\xi}}d\theta\nonumber\\
&\;\;\;\;+\frac{1}{i2\pi}\int_{-\ep}^{0}\frac{|z|^\al|z-s|^\la}{z-\xi}dz.
\end{align}
After some calculations, and set $\ep=2s,$ one has
\bq\label{RE84}
f(\xi)=\frac{\xi^{\al+\la}}{i2\sin{(\al+\la)\pi}}\left(1+\mathcal{O}\left(\frac{s}{\va_{1}}\right)+\mathcal{O}\left(\left(\frac{s}{\va_{1}}\right)^{\al+\la+1}\right)\right), \;\;2s=\ep\ll\left|\xi\right|=\va_{1}\leq \va,
\eq
valid for $\al+\la+1>0,$ where we have used the following estimations,
\bbq\label{RE85}
\left|\lim_{\va'\rw0}\int_{-\pi+\va'}^{\pi-\va'}
\frac{(\ep{e^{i\theta}})^{\al+1}(\ep{e^{i\theta}}-s)^{\la}}{\ep{e^{i\theta}-\xi}}d\theta\right|\leq\frac{2\pi\ep^{\al+1}(\ep+s)^{\la}}{\va_{1}-\ep},
\eeq
\bbq\label{RE86}
\left|\int_{-\ep}^{0}\frac{|z|^\al||z-s|^\la}{z-\xi}dz\right|\leq\frac{\ep^{\al+1}(\ep+s)^{\la}}{\va_{1}-\ep}.
\eeq

\begin{center}
\begin{tikzpicture}
\begin{scope}[line width=2pt]
\draw[-, >=latex] ({0}:3.8)
    arc ({0}:{175}:3.8);
\draw[-, >=latex] ({0}:3.8)
    arc ({0}:{-175}:3.8);
\draw[-] (-3.8,.288)--(0,.288);
\draw[-] (-3.8,-.324)--(0,-.324);
\draw[dashed, >=latex] ({0}:1.5)
    arc ({0}:{168}:1.5);
\draw[dashed, >=latex] ({0}:1.5)
    arc ({0}:{-168}:1.5);
\draw[dashdotted, >=latex] ({0}:2.85)
    arc ({0}:{175}:2.85);
\draw[dashdotted, >=latex] ({0}:2.85)
    arc ({0}:{-175}:2.85);

\fill (0,0) circle (1pt);

\draw[dashdotted] (-2.85,0)--(0,0);

\draw[->,>=stealth] (-3.8,.288)--(-2,.288);

\draw[<-,>=stealth] (-3.3,-.324)--(-2,-.324);

\draw[<-,>=stealth] (0,2.85)--(0.01,2.85);
\draw[<-,>=stealth] (0,-2.85)--(-0.01,-2.85);

\draw[->,>=stealth] (-0.937,-1.159)--(-0.956,-1.144);
\draw[<-,>=stealth] (-0.937,1.159)--(-0.956,1.144);

\draw[<-,>=stealth] (-2.374,-2.936)--(-2.422,-2.898);
\draw[->,>=stealth] (-2.374,2.936)--(-2.422,2.898);

\draw[dotted] (0,0)--(-3.29,1.9);
\node at (-2.9,1.5) {$\va_{2}$};

\draw[dotted] (0,0)--(0,2.85);
\node at (-.11,2.2) { $\va$};
\draw[dotted] (0,0)--(.75,-1.294);
\node at (.4,-1.) { $\epsilon$};
\node at (-0.1,-0.15) {$o$};
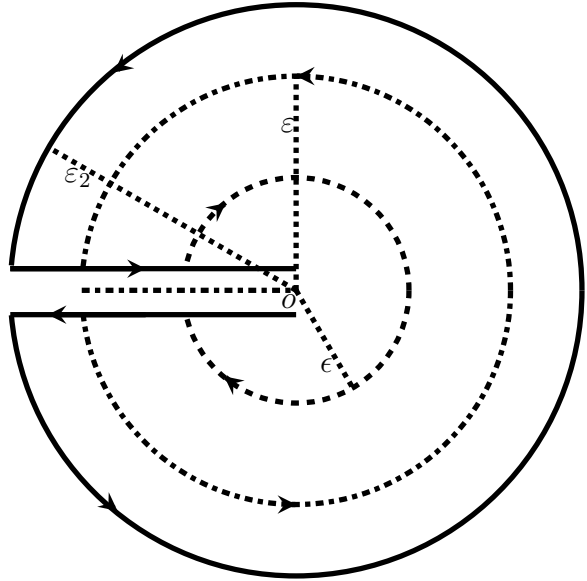
\node[below] at (0,-4) {Figure 6. The Integration Contours};
\end{scope}
\end{tikzpicture}
\end{center}

\begin{rem}
In order to compare the results of the situation of the singularly perturbed Laguerre weight in \cite{XDZ2014}, we take the same integration contour of (5.11) therein as the integration contour $\Gamma$ in (\ref{RE82}).
They are different, for example the lower bound $\delta$ satisfies $s/\delta=\mathcal{O}(1),$ and takes $\delta=s$ in \cite{XDZ2014}, but in our case the lower bound $\ep>s$ and set $\ep=2s.$

It is interesting to investigate the behavior of $\Phi_{1}(\xi,s)$ on the boundary $\partial U(0,\va).$
Substituting (\ref{RE84}), into (\ref{RE78}) and set $\va_{1}=\va,$ then one finds
\bq\label{RE87}
\Phi_{1}(\xi,s)J^{-1}=G(\xi)\left(
\begin{matrix}
1&\frac{\xi^{\al+\la}}{i2\sin{(\al+\la)\pi}}\left(1+\mathcal{O}\left(\frac{s}{\va}\right)+\mathcal{O}\left(\left(\frac{s}{\va}\right)^{\al+\la+1}\right)\right)\\
0&1
\end{matrix}
\right)
\xi^{\frac{\al+\la}{2}\sigma_{3}}\left(1-\frac{s}{\xi}\right)^{\frac{\la}{2}\sigma_{3}},
\eq
if $s=0,$ the right hand side of $(\ref{RE87})$ is the same as (\ref{RE76}).
\par
From (\ref{RE78}), (\ref{RE84}) and $\Phi_{0}(\xi)$ in (\ref{RE57}), then one follows the matching condition (\ref{RE80}).
\end{rem}
\par
For $\al+\la\in \mathbb{N},$ then the modified Bessel matrix function has a factorization as follows
\bq\label{RE88}
\left(
\begin{matrix}
I_{\alpha+\lambda}\left(\xi^{\frac{1}{2}}\right)&i\frac{1}{\pi}K_{\alpha+\lambda}\left(\xi^{\frac{1}{2}}\right)\\
i\pi\xi^{\frac{1}{2}}I'_{\alpha+\lambda}\left(\xi^{\frac{1}{2}}\right)&-\xi^{\frac{1}{2}}K'_{\alpha+\lambda}\left(\xi^{\frac{1}{2}}\right)
\end{matrix}
\right)
=\tilde{G}(\xi)\xi^{\frac{\alpha+\lambda}{2}\sigma_{3}}\left(
\begin{matrix}
1&\frac{i(-1)^{\al+\la+1}}{\pi}\log\frac{\sqrt{\xi}}{2}\\
0&1
\end{matrix}
\right),
\eq
where
\bbq\label{RE89}
\tilde{G}(\xi)=\left(
\begin{matrix}
\xi^{-\frac{\alpha+\lambda}{2}}I_{\alpha+\lambda}\left(\xi^{\frac{1}{2}}\right)&\frac{i(-1)^{\al+\la}}{2\pi}\xi^{\frac{\alpha+\lambda}{2}}I_{-(\alpha+\lambda)}\left(\xi^{\frac{1}{2}}\right)\\
i\pi\xi^{\frac{1-(\alpha+\lambda)}{2}}I'_{\alpha+\lambda}\left(\xi^{\frac{1}{2}}\right)&\frac{(-1)^{\al+\la+1}}{2}\xi^{\frac{1+\alpha+\lambda}{2}}I'_{-(\alpha+\lambda)}\left(\xi^{\frac{1}{2}}\right)
\end{matrix}
\right),
\eeq
and $\tilde{G}(\xi)$ is an entire matrix function. One verifies with the following formulas,
\bbq\label{RE62}
K_{\up}=\frac{\pi}{2}\lim_{\up\rw m}\frac{I_{-\up}(z)-I_{\up}(z)}{\sin{\up\pi}}=\frac{(-1)^{m}}{2}\left(\left.\frac{\partial}{\partial{\up}}I_{\up}(z)\right|_{\up=-m}-\left.\frac{\partial}{\partial{\up}}I_{\up}(z)\right|_{\up=m}\right),
\;\;\up=m \in \mathbb{N},
\eeq
where
\begin{align*}\label{RE63}
&I_{-m}\left(z\right)=\left.\frac{\partial}{\partial{\up}}I_{\up}(z)\right|_{\up=-m}\nonumber\\
&=\left(\frac{z}{2}\right)^{-m}\sum_{j=0}^{m-1}\frac{(-1)^{m-j}(m-j-1)!}{\Gamma(j+1)}\left(\frac{z^{2}}{4}\right)^{j}
+\left(\frac{z}{2}\right)^{m}\sum_{j=0}^{\infty}\frac{-\log\frac{z}{2}+\psi(j+1)}{\Gamma(j+1)\Gamma(j+m+1)}\left(\frac{z^{2}}{4}\right)^{j},
\end{align*}
\bbq\label{RE64}
\left.\frac{\partial}{\partial{\up}}I_{\up}(z)\right|_{\up=m}=\left(\frac{z}{2}\right)^{m}\sum_{j=0}^{\infty}\frac{\log{\frac{z}{2}}-\psi(j+m+1)}{\Gamma(j+1)\Gamma(j+m+1)}\left(\frac{z^{2}}{4}\right)^{j},
\eeq
and
\bbq\label{RE65}
\psi(x)=\frac{\Gamma'(x)}{\Gamma(x)}.
\eeq
\par
With $f(\xi)$ in (\ref{RE78}) replaced by $\tilde{f}(\xi),$ which satisfies $(\ref{RE81}),$ then one can define
\bq\label{RE90}
\tilde{f}(\xi)=\frac{(-1)^{\al+\la+1}}{2\pi^{2}}\int_{\Gamma}\frac{z^{\al}(z-s)^{\la}\log\left(\frac{\sqrt{z}}{2}\right)}{z-\xi}dz,\;\;  \xi \in U(0,\va)\setminus (-\va,0),\;\; \arg\xi \in (-\pi, \pi),
\eq
where the integration contour $\Gamma$ keeps the same as in (\ref{RE82}).
\par
For $\xi\in \partial U(0,\va)$ and $s\rw 0,$ the matching condition (\ref{T11}) can be verified with the similar steps of
(\ref{RE83})-(\ref{RE84}), one gets
\bq\label{RE91}
\tilde{f}(\xi)=\frac{(-1)^{\al+\la}}{i\pi}\xi^{\al+\la}\log{\frac{\sqrt{\xi}}{2}}\left(1+\mathcal{O}\left(\frac{s}{\va_{1}}\right)+\mathcal{O}\left(\left(\frac{s}{\va_{1}}\right)^{\al+\la+1}\log{s}\right)\right),\;\;
\eq
where $2s=\ep\ll\left|\xi\right|=\va_{1}\leq \va.$
\par
At last, we define
\bq\label{RE92}
R_{0}(\xi,s)=\left\{
\begin{array}{ll}
\Phi(\xi,s)\Phi_{0}^{-1}(\xi), & \xi \in \mathbb{C}\setminus U(0,\va),\\
\\
\Phi(\xi,s)\Phi_{1}^{-1}(\xi, s),& \xi \in U(0,\va).\\
\end{array}
\right.
\eq
Then the matrix function $R_{0}(\xi, s)$ fulfills the following properties,
\par
$(a)$ $R_{0}(\xi, s)$ is analytic in $\xi \in \mathbb{C}\setminus \partial U(0,\va).$
\par
$(b)$ From (\ref{RE78}) and (\ref{RE91}), then $R_{0}(\xi, s)$ satisfies the following jump on the counter clockwise circle $\partial U(0,\va),$
\bq\label{RE93}
(R_{0})_{+}(\xi,s)=(R_{0})_{-}(\xi,s)\left\{
\begin{array}{ll}
I+\mathcal{O}\left(h_{1}(s)\right), & \al+\la\notin \mathbb{N},\\
\\
I+\mathcal{O}\left(h_{2}(s)\right), & \al+\la\in \mathbb{N},\\
\end{array}
\right.
\eq
where
\bq\label{R95}
h_{1}(s)=\left\{
\begin{array}{ll}
s,& \al+\la>0,\\
\\
s^{\al+\la+1}, & -1<\al+\la<0,\\
\end{array}
\right.
\;\;{\rm and}\;\;
h_{2}(s)=\left\{
\begin{array}{ll}
s,& \al+\la>0,\\
\\
s^{\al+\la+1}\log{s}, & -1<\al+\la<0.\\
\end{array}
\right.
\eq
\par
With a similar argument as (\ref{Rw07}), (\ref{Rw08}), (\ref{Rw09}) and (\ref{Rw10}) in section 3.7, we have
\par
$(c)$ The uniformly asymptotic behavior of $R_{0}(\xi, s)$ for bounded $\xi,$ $s<\ep\ll\left|\xi\right|=\va_{1}\leq \va$ and $s\rw 0,$
\bq\label{RE94}
R_{0}(\xi,s)=\left\{
\begin{array}{ll}
I+\mathcal{O}(h_{1}(s)), & \al+\la\notin \mathbb{N},\\
\\
I+\mathcal{O}(h_{2}(s)), & \al+\la\in \mathbb{N},\\
\end{array}
\right.
 \eq
 where $h_{1}(s)$ and $h_{2}(s)$ are given by $(\ref{R95}).$
 \par
 $(d)$ The uniformly asymptotic behavior of $R_{0}(\xi, s)$ for $\xi\rw \infty$ and $s\rw 0,$
\bq\label{RE95}
R_{0}(\xi,s)=\left\{
\begin{array}{ll}
I+\mathcal{O}\left(h_{1}(s)\xi^{-1}\right), &\al+\la\notin \mathbb{N},\\
\\
I+\mathcal{O}\left(h_{2}(s)\xi^{-1}\right), &\al+\la\in \mathbb{N}.\\
\end{array}
\right.
\eq
It completes the nonlinear steepest descent analysis for $\Phi(\xi,s)$ as $s\rw 0.$
\\
\par
{\bf Proof of Theorem 2}
\begin{proof}
Combining (\ref{Rw13})-(\ref{RR01}), (\ref{RE59}), (\ref{RE94})-(\ref{RE95}) and set $\xi=n^{2}\phi^{2}(z)$ in (\ref{RE59}) for $\xi\in \widehat{\Omega}_{3},$ then
\begin{align}\label{Rw23}
\left(
\begin{matrix}
\varphi_{1}\left(\xi,s\right)\\
\varphi_{2}\left(\xi,s\right)\\
\end{matrix}
\right)
&=\Phi_{-}(\xi,s)e^{\frac{i\pi(\al+\la-1)}{2}\sigma_{3}}
\left(
\begin{matrix}
1\\
1\\
\end{matrix}
\right)\nonumber\\
&=R_{0}(\xi,s)M_{0}\left(
\begin{matrix}
I_{\alpha+\lambda}\left(\xi^{\frac{1}{2}}\right)&i\frac{1}{\pi}K_{\alpha+\lambda}\left(\xi^{\frac{1}{2}}\right)\\
i\pi\xi^{\frac{1}{2}}I'_{\alpha+\lambda}\left(\xi^{\frac{1}{2}}\right)&-\xi^{\frac{1}{2}}K'_{\alpha+\lambda}\left(\xi^{\frac{1}{2}}\right)
\end{matrix}
\right)\left(
\begin{matrix}
e^{\frac{i\pi}{2}(\lambda+\alpha-1)}\\
0\\
\end{matrix}
\right)\nonumber\\
&=R_{0}(\xi,s)M_{0}\left(
\begin{matrix}
-iJ_{\alpha+\lambda}\left(|\xi|^{\frac{1}{2}}\right)\\
\pi|\xi|^{\frac{1}{2}}J'_{\alpha+\lambda}\left(|\xi|^{\frac{1}{2}}\right)
\end{matrix}
\right),
\end{align}
where $M_{0}$ and $R_{0}(\xi,s)$ are given by (\ref{RE60}) and (\ref{RE94}), respectively, and it has been used the facts that for $\arg{z}\in (-\pi,\pi/2],$ $e^{\frac{i\pi\beta}{2}}I_{\beta}(ze^{-\frac{i\pi}{2}})=J_{\beta}(z),$ see \cite{OLC2010}.
\par
Let $\xi=n^{2}\phi^{2}(x),$ $x=\frac{u}{16n^{2}},$ then
\begin{align}\label{Rw24}
\left(
\begin{matrix}
\varphi_{1}\left(n^{2}\phi^{2}(x),s\right)\\
\varphi_{2}\left(n^{2}\phi^{2}(x),s\right)\\
\end{matrix}
\right)
=\left\{
\begin{array}{ll}
\left(I+\mathcal{O}(h_{1}(s))\right)M_{0}\left(
\begin{matrix}
-iJ_{\alpha+\lambda}\left(u\right)\\
\pi|\xi|^{\frac{1}{2}}J'_{\alpha+\lambda}\left(u\right)
\end{matrix}
\right)+\mathcal{O}\left(\frac{1}{n^{2}}\right), & \al+\la\notin \mathbb{N},\\
\\
\left(I+\mathcal{O}(h_{2}(s))\right)M_{0}\left(
\begin{matrix}
-iJ_{\alpha+\lambda}\left(u\right)\\
\pi|\xi|^{\frac{1}{2}}J'_{\alpha+\lambda}\left(u\right)
\end{matrix}
\right)+\mathcal{O}\left(\frac{1}{n^{2}}\right), & \al+\la\in \mathbb{N},\\
\end{array}
\right.
\end{align}
where $h_{1}(s)$ and $h_{2}(s)$ are given by $(\ref{R95}).$ 
Moreover, inserting (\ref{Rw24}) and similarly estimations for $\xi=n^{2}\phi^{2}(y)$ and $y=\frac{{\rm v}}{16n^{2}}$ into (\ref{RR01}), then
\begin{align}\label{Rw25}
\frac{1}{4n}K_{n}(\frac{u}{4n},&\frac{{\rm v}}{4n}; s)=\frac{\varphi_{1}(-{\rm v},s)\varphi_{2}(-u,s)-\varphi_{1}(-u,s)\varphi_{2}(-{\rm v},s)}{i2\pi(u-{\rm v})}+\mathcal{O}\left(\frac{1}{n^{2}}\right)\nonumber\\
&=\left\{
\begin{array}{ll}
\frac{J_{\al+\la}(u){\rm v}J'_{\al+\la}({\rm v})-J_{\al+\la}({\rm v})uJ'_{\al+\la}(u)}{2(u-{\rm v})}+\mathcal{O}(h_{1}(s))+\mathcal{O}\left(\frac{1}{n^{2}}\right), & \al+\la\notin \mathbb{N},\\
\\
\frac{J_{\al+\la}(u){\rm v}J'_{\al+\la}({\rm v})-J_{\al+\la}({\rm v})uJ'_{\al+\la}(u)}{2(u-{\rm v})}+\mathcal{O}(h_{2}(s))+\mathcal{O}\left(\frac{1}{n^{2}}\right), & \al+\la\in \mathbb{N}.\\
\end{array}
\right.
\end{align}
For $n\rw \infty,$ (\ref{Rw25}) follows (\ref{T8}).
\end{proof}

\subsection{Painlev\'e \uppercase\expandafter{\romannumeral5} kernel to the Bessel kernel $\mathbf{J}_{\al}$ as $s\rightarrow \infty$}
\par
We start with the RH problem (\ref{RE20})-(\ref{RE23}) for $\Phi(\xi,s),$ let $\xi=s\hat{z},$ and define
\bq\label{RE96}
A(\hat{z},s):=s^{\frac{1}{4}\sigma_{3}}\Phi(s\hat{z},s)e^{-\sqrt{s\hat{z}}\sigma_{3}},
\eq
then $A(\hat{z},s)$ satisfies the following RH problem.
\par
$(a)$ $A(\hat{z},s)$ is analytic in $\mathbb{C}\setminus \displaystyle\cup_{j=1}^{3}\Sigma'_{j}\cup\left(0,1\right),$ see Figure 7.
\par
$(b)$ $A(\hat{z},s)$ fulfills the jump relation and $\arg{\hat{z}} \in (-\pi, \pi),$
\bq\label{RE97}
A_{+}(\hat{z},s)=A_{-}(\hat{z},s)\left\{
\begin{array}{llll}
e^{{i}\lambda\pi{\sigma_{3}}}, & \hat{z}\in \left(0,1\right), \\
\\
\left(
\begin{matrix}
1&0\\
e^{i(\lambda+\alpha)\pi-2\sqrt{s\hat{z}}}&1\\
\end{matrix}
\right),& \hat{z}\in \Sigma'_{1},\\
\\
\left(
\begin{matrix}
0&1\\
-1&0\\
\end{matrix}
\right),& \hat{z} \in \Sigma'_{2},\\
\\
\left(
\begin{matrix}
1&0\\
e^{-i(\lambda+\alpha)\pi-2\sqrt{s\hat{z}}}&1\\
\end{matrix}
\right),& \hat{z} \in \Sigma'_{3}.
\end{array}
\right.
\eq
\par
$(c)$ As $\hat{z}\rightarrow \infty,$
\bq\label{RE98}
A(\hat{z},s)=\left(I+\mathcal{O}\left(\frac{1}{\hat{z}}\right)\right)\hat{z}^{-\frac{1}{4}\sigma_{3}}\frac{I+i\sigma_{1}}{\sqrt{2}},\;\;\arg{\hat{z}} \in (-\pi, \pi).
\eq
\par
$(d)$ As $\hat{z}\rightarrow 0,$
\bbq\label{RE99}
A(z,s)=Q_{3}(s)\left(I+\mathcal{O}(\hat{z})\right)\hat{z}^{\frac{\alpha}{2}\sigma_{3}},\;\;\arg{\hat{z}} \in (-\pi, \pi).
\eeq
\par
$(e)$ As $\hat{z}\rw 1,$
\bbq\label{RE100}
A(\hat{z},s)=\mathcal{O}(1)(\hat{z}-1)^{\frac{\la}{2}\sigma_{3}},\;\;\arg{(\hat{z}-1)} \in (-\pi, \pi).
\eeq

\begin{center}
\begin{tikzpicture}
\begin{scope}[line width=2pt]
\draw[->,>=stealth] (0,0)--(1.5,0);
\draw[-] (0,0)--(2.5,0);
\node[below] at (2.5,0) {$1$};

\draw[->,>=stealth] (-5,0)--(-2,0);
\draw[-] (-2.8,0)--(0,0);
\draw[dashed] (2.5,0)--(5.6,0);
\draw[->,>=stealth] (-3.84,3.36)--(-1.92,1.68);
\draw[-] (-3.84,3.36)--(0,0);
\draw[->,>=stealth] (-3.84,-3.36)--(-1.92,-1.68);
\draw[-] (-3.84,-3.36)--(0,0);
\node[below] at (0,0) {$O$};
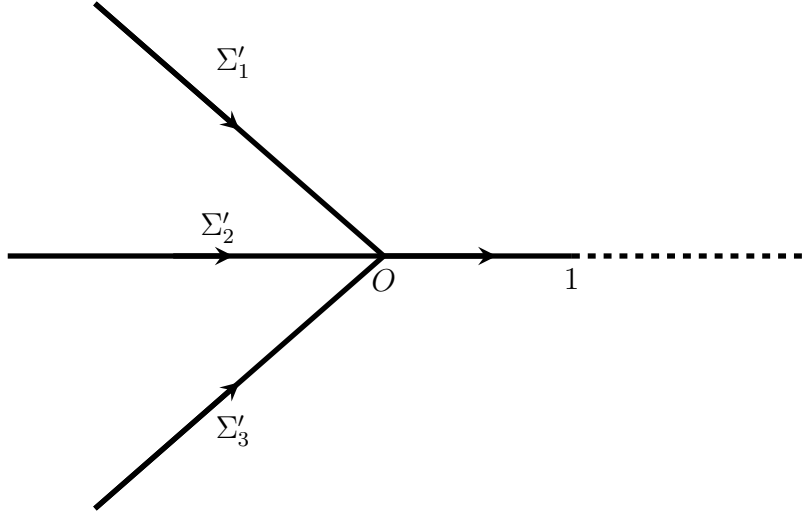
\node[below] at (-.8,-4) { Fig.7. Contours of $ \displaystyle\cup_{j=1}^{3}\Sigma'_{j}\cup\left(0,1\right).$}; 

\node[above] at (-2,-2.7) {{$\Sigma'_{3}$}};
\node[above] at (-2.2,0) {{$\Sigma'_{2}$}};
\node[above] at (-2,2.2) {{$\Sigma'_{1}$}};
\end{scope}
\end{tikzpicture}
\end{center}

\par
From the jump condition of (\ref{RE97}), as $s \rw \infty$ and $\hat{z}$ away from the origin, then the jumps on $\Sigma'_{1}$ and $\Sigma'_{3}$ tend to identical matrices. So, $A(\hat{z},s)$ can be approximated by $B(\hat{z})$ which independents of $s,$ and $B(\hat{z})$ satisfies the following RH problem.
\par
$(a)$ $B(\hat{z})$ is analytic in $\mathbb{C}\setminus (-\infty, 1).$
\par
$(b)$ $B(\hat{z})$ satisfies the jump condition
\bbq\label{RE101}
B_{+}(\hat{z})=B_{-}(\hat{z})\left\{
\begin{array}{llll}
e^{{i}\lambda\pi{\sigma_{3}}}, & \hat{z}\in \left(0,1\right), \\
\\
\left(
\begin{matrix}
0&1\\
-1&0\\
\end{matrix}
\right),& \hat{z} \in (-\infty, 0),\\
\end{array}
\right.
\eeq
where $\arg \xi \in (-\pi,\pi).$
\par
$(c)$ For $\hat{z} \rw \infty,$
\bbq\label{RE102}
B(\hat{z})=\left(I+\mathcal{O}\left(\frac{1}{\hat{z}}\right)\right)\hat{z}^{-\frac{1}{4}\sigma_{3}}\frac{I+i\sigma_{1}}{\sqrt{2}},\;\;\arg\xi\in(-\pi,\pi).
\eeq
\par
$(d)$ For $\hat{z}=1,$
\bbq\label{RE103}
B(\hat{z})=\mathcal{O}(1)(\hat{z}-1)^{\frac{\la}{2}\sigma_{3}},\;\;\arg\xi\in(-\pi,\pi).
\eeq
\par
By the Szeg\"{o} function, $B(\hat{z})$ can be constructed as follows
\begin{align}\label{RE104}
B(\hat{z})&=\left(
\begin{matrix}
1&0\\
i\la&1\\
\end{matrix}
\right)
\hat{z}^{-\frac{1}{4}\sigma_{3}}
\frac{I+i\sigma_{1}}{\sqrt{2}}
\left(\exp\left(\frac{\la{\sqrt{\hat{z}}}}{2\pi}\int_{-\infty}^{0}\frac{\log{\frac{x-1}{x}}}{\sqrt{-x}}\frac{1}{x-\hat{z}}dx\right)\right)^{\sigma_{3}}\left(\frac{\hat{z}-1}{\hat{z}}\right)^{\frac{\la}{2}\sigma_{3}}\nonumber\\
&=\left(
\begin{matrix}
1&0\\
i\la&1\\
\end{matrix}
\right)\hat{z}^{-\frac{1}{4}\sigma_{3}}
\frac{I+i\sigma_{1}}{\sqrt{2}}\left(\exp\left(\frac{\la\sqrt{\hat{z}}}{2}\int_{0}^{1}\frac{1}{\sqrt{x}}\frac{1}{x-\hat{z}}dx\right)\right)^{\sigma_{3}}\nonumber\\
&=\left(
\begin{matrix}
1&0\\
i\la&1\\
\end{matrix}
\right)\hat{z}^{-\frac{1}{4}\sigma_{3}}\frac{I+i\sigma_{1}}{\sqrt{2}}\left(\frac{\hat{z}-1}{(\sqrt{\hat{z}}+1)^{2}}\right)^{\frac{\la}{2}\sigma_{3}}, \;\;\arg{\hat{z}}\in (-\pi,\pi).
\end{align}
\par

\begin{rem}
From (\ref{RE21}), (\ref{RE34}) and (\ref{RE96}), it follows that
\bq\label{R4}
A(\hat{z},s)\frac{I+i\sigma_{1}}{\sqrt{2}}\hat{z}^{\frac{1}{4}\sigma_{3}}=I+\left(
\begin{matrix}
q(s)s^{-1}&-ir(s)s^{\frac{1}{2}}\\
it(s)s^{-\frac{3}{2}}&-q(s)s^{-1}\\
\end{matrix}
\right)\frac{1}{\hat{z}}+\mathcal{O}\left(\frac{1}{\hat{z}^{2}}\right).
\eq
By $A(\hat{z},s)$ can be approximated by $B(\hat{z})$ for large $s$ and (\ref{RE104}), then one finds,
\bq\label{R5}
B(\hat{z})\frac{I+i\sigma_{1}}{\sqrt{2}}\hat{z}^{\frac{1}{4}\sigma_{3}}=I+\left(
\begin{matrix}
\la^{2}/2&i\la\\
i\la(\la^{2}-1)/3&-\la^{2}/2\\
\end{matrix}
\right)\frac{1}{\hat{z}}+\mathcal{O}\left(\frac{1}{\hat{z}^{2}}\right).
\eq
From (\ref{R4}) and (\ref{R5}), one obtains the initial data for $q(s),$ $r(s)$ and $t(s),$
\bq\label{R10}
r(s)=-\la{s^{-\frac{1}{2}}}+\mathcal{O}\left(s^{-1}\right),\;\;q(s)=\frac{\la^{2}}{2}s+\mathcal{O}\left(s^{\frac{1}{2}}\right),\;\;t(s)=\frac{\la(\la^{2}-1)}{3}s^{\frac{3}{2}}+\mathcal{O}\left(s\right),\;\;s\rw \infty.
\eq
\end{rem}

As $\hat{z} \rw 0,$ $B(\hat{z})$ can not be applied to approximate $A(\hat{z},s),$ and the jumps on $\Sigma'_{1}$ and $\Sigma'_{3}$ in (\ref{RE97}) may have oscillation entries. Hence, it is need to construct a local parametrix $B_{0}(\hat{z},s)$ in the disk $\left|\hat{z}\right|<1,$ which satisfies the jump in (\ref{RE97}) of $A(\hat{z},s),$ and matches the following condition.
\bq\label{RE105}
B(\hat{z})B_{0}^{-1}(\hat{z},s)=I+\mathcal{O}\left(s^{-\frac{1}{2}}\right),\;\; \left|\hat{z}\right|=1,\;\;s \rw \infty.
\eq
\par
We construct the matrix function $B_{0}(\hat{z},s)$ as follows
\bq\label{RE106}
B_{0}(\hat{z},s)=E_{0L}(\hat{z})F(\hat{z},s)\left\{
\begin{array}{llll}
e^{-\frac{\sqrt{s\hat{z}}-i\la\pi}{2}\sigma_{3}}, & 0<\arg \hat{z}<\pi, \\
\\
e^{-\frac{\sqrt{s\hat{z}}+i\la\pi}{2}\sigma_{3}},& -\pi<\arg \hat{z}<0,\\
\end{array}
\right.
\eq
where $E_{0L}(\hat{z})$ is analytic in $U(0,1).$ $F(\hat{z},s)$ fulfills the following RH problem.
\par
$(a)$ $F(\hat{z},s)$ is analytic in $\mathbb{C}\setminus \bigcup_{j=I}^{III}\Sigma_{j},$ see Fig. 8.
\par
$(b)$ $F(\hat{z},s)$ satisfies the jump condition as follows
\bbq\label{RE107}
F_{+}(\hat{z},s)=F_{-}(\hat{z},s)\left\{
\begin{array}{llll}
\left(
\begin{matrix}
1&0\\
e^{i\al\pi}&1\\
\end{matrix}
\right),& \hat{z}\in \Sigma_{I},\\
\\
\left(
\begin{matrix}
0&1\\
-1&0\\
\end{matrix}
\right),& \hat{z} \in \Sigma_{II},\\
\\
\left(
\begin{matrix}
1&0\\
e^{-i\al\pi}&1\\
\end{matrix}
\right),& \hat{z} \in \Sigma_{{\tiny III}},
\end{array}
\right.
\eeq
where $\arg \hat{z} \in (-\pi, \pi).$
\par
$(c)$ As $\hat{z} \rw \infty,$
\bbq\label{RE108}
F(\hat{z},s)=\left(I+\mathcal{O}\left(\frac{1}{\hat{z}}\right)\right)\hat{z}^{-\frac{1}{4}\sigma_{3}}\frac{I+i\sigma_{1}}{\sqrt{2}}e^{\sqrt{s\hat{z}}\sigma_{3}},\;\; \arg \hat{z} \in (-\pi, \pi).
\eeq
\par
$(d)$ As $\hat{z} \rw 0,$
\bbq\label{RE109}
F(\hat{z}, s)=\mathcal{O}(1)\hat{z}^{\frac{\al}{2}\sigma_{3}},\;\; \arg \hat{z} \in (-\pi, \pi).
\eeq
\par

\begin{center}
\begin{tikzpicture}
\begin{scope}[line width=2pt]

\draw[->,>=stealth] (-3,0)--(-2,0);
\draw[-] (-5.6,0)--(0,0);
\draw[->,>=stealth] (-3.84,3.36)--(-1.92,1.68);
\draw[-] (-3.84,3.36)--(0,0);
\draw[->,>=stealth] (-3.84,-3.36)--(-1.92,-1.68);
\draw[-] (-3.84,-3.36)--(0,0);
\node[below] at (0,0) {$O$};
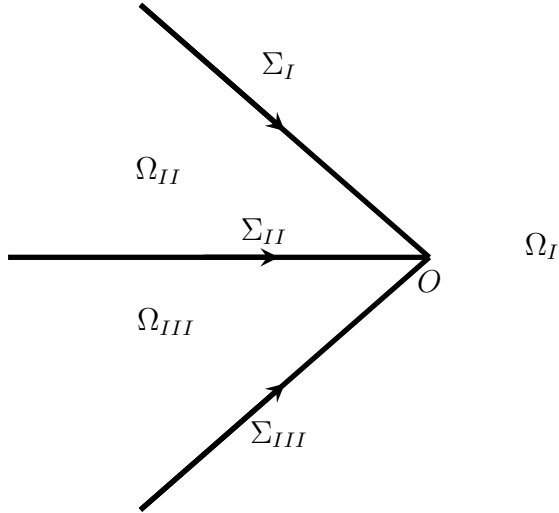
\node[below] at (-.8,-4) { Figure 8. Contours $\displaystyle\cup_{j=I}^{III}\Sigma_{j},$ and regions $\Omega_{j},$ $j=I, II, III.$};

\node[above] at (-2,-2.7) {{$\Sigma_{III}$}};
\node[above] at (-2.2,0) {{$\Sigma_{II}$}};
\node[above] at (-2,2.2) {{$\Sigma_{I}$}};

\node[above] at (1.5,-.2) {{$\Omega_{I}$}};
\node[above] at (-3.6,0.8) {{$\Omega_{II}$}};
\node[above] at (-3.5,-1.2) {{$\Omega_{III}$}};
\end{scope}
\end{tikzpicture}
\end{center}

\par
By the modified Bessel functions, $F(\hat{z},s)$ can be constructed as follows
\bq\label{RE110}
F(\hat{z},s)=
M_{1}\left\{
\begin{array}{llll}
\left(
\begin{matrix}
I_{\alpha}\left(\sqrt{s\hat{z}}\right)&i\frac{1}{\pi}K_{\alpha}\left(\sqrt{s\hat{z}}\right)\\
i\pi\sqrt{s\hat{z}}I'_{\alpha}\left(\sqrt{s\hat{z}}\right)&-\sqrt{s\hat{z}}K'_{\alpha}\left(\sqrt{s\hat{z}}\right)
\end{matrix}
\right), &\hat{z} \in \Omega_{I},\\
\\
\left(
\begin{matrix}
I_{\alpha}\left(\sqrt{s\hat{z}}\right)&i\frac{1}{\pi}K_{\alpha}\left(\sqrt{s\hat{z}}\right)\\
i\pi\sqrt{s\hat{z}}I'_{\alpha}\left(\sqrt{s\hat{z}}\right)&-\sqrt{s\hat{z}}K'_{\alpha}\left(\sqrt{s\hat{z}}\right)
\end{matrix}
\right)\left(
\begin{matrix}
1&0\\
-e^{i\pi\al}&1\\
\end{matrix}
\right),& \hat{z} \in \Omega_{II},\\
\\
\left(
\begin{matrix}
I_{\alpha}\left(\sqrt{s\hat{z}}\right)&i\frac{1}{\pi}K_{\alpha}\left(\sqrt{s\hat{z}}\right)\\
i\pi\sqrt{s\hat{z}}I'_{\alpha}\left(\sqrt{s\hat{z}}\right)&-\sqrt{s\hat{z}}K'_{\alpha}\left(\sqrt{s\hat{z}}\right)
\end{matrix}
\right)\left(
\begin{matrix}
1&0\\
e^{-i\pi\al}&1\\
\end{matrix}
\right),& \hat{z} \in \Omega_{III},
\end{array}
\right.
\eq
where the regions $\Omega_{j}, j=I, II, III,$ illustrated in Fig.8, $M_{1}$ only dependents on $s,$
\bq\label{Ro2}
M_{1}=\left(i\left(4\al^{2}+3\right)s^{-\frac{1}{2}}\sigma_{-}/8+I\right)s^{\frac{1}{4}\sigma_{3}}\pi^{\frac{1}{2}\sigma_{3}}.
\eq
\par
With the matching condition $B_{0}(\hat{z},s)\sim B(\hat{z})$ for $\hat{z} \in \partial U(0,1)$ and the expression of  $B(\hat{z})$ in (\ref{RE104}), then the matrix function $E_{0L}(\hat{z})$ is given by
\bq\label{RE111}
E_{0L}(\hat{z})=B(\hat{z})\left\{
\begin{array}{llll}
e^{-\frac{i\la\pi}{2}\sigma_{3}}\frac{I-i\sigma_{1}}{\sqrt{2}}\hat{z}^{-\frac{1}{4}\sigma_{3}}, & \arg \hat{z} \in (0, \pi), \\
\\
e^{\frac{i\la\pi}{2}\sigma_{3}}\frac{I-i\sigma_{1}}{\sqrt{2}}\hat{z}^{-\frac{1}{4}\sigma_{3}},& \arg \hat{z} \in (-\pi,0).\\
\end{array}
\right.
\eq
One checks that $E_{0L}(\hat{z})$ has no jump on the real axis and is analytic in the unite disk. $E_{0L}(\hat{z})$ has square root singularities at $\hat{z}=0,$ hence, these singularities are weak and removable.
\par
From (\ref{RE106}), (\ref{RE110}) and (\ref{RE111}), it follows the matching condition (\ref{RE105}) which is uniform for bounded $\hat{z}.$
\par
We define
\bq\label{RE112}
R_{0L}(\hat{z},s)=\left\{
\begin{array}{ll}
A(\hat{z},s)B^{-1}(\hat{z}), & \left|\hat{z}\right|>1,\\
\\
A(\hat{z},s)B_{0}^{-1}(\hat{z}, s),& \left|\hat{z}\right|<1.\\
\end{array}
\right.
\eq
$R_{0L}(\hat{z},s)$ is a piecewise analytic function in $\mathbb{C}\setminus \Sigma_{R_{0L}},$ where the contour $\Sigma_{R_{0L}}$ contains the counter clockwise unite circle, parts of $\Sigma_{I}$ and $\Sigma_{III}$ (see Figure 8) for $\left|\hat{z}\right|>1,$ then one has
\bq\label{RE113}
J_{R_{0L}}(\hat{z},s)=\left\{
\begin{array}{ll}
I+\mathcal{O}\left(s^{-\frac{1}{2}}\right), & \left|\hat{z}\right|=1,\\
\\
I+\mathcal{O}\left(e^{-c\sqrt{s}}\right),& \hat{z}\in \{\hat{z}\in \Sigma_{I}\cap\Sigma_{III}: \left|\hat{z}\right|>1\},\\
\end{array}
\right.
\eq
where $c$ is a positive constant.
\par
With a similar argument in section 3.7, one finds,
\bq\label{RE114}
R_{0L}(\hat{z},s)=I+\mathcal{O}(s^{-\frac{1}{2}}),
\eq
where the error term is uniform for bounded $\hat{z}\in (0,\infty)$ and $t\in(0,c],$ $c$ is a finite and positive constant.
\\
\par
{\bfseries Proof of Theorem 3}
\begin{proof}
With the help of (\ref{Rw13}), (\ref{RE96}), (\ref{RE110}) and (\ref{RE114}), one finds,
\begin{align}\label{Ro1}
\left(
\begin{matrix}
\varphi_{1}\left(\xi,s\right)\\
\varphi_{2}\left(\xi,s\right)\\
\end{matrix}
\right)
&=\Phi_{-}(\xi,s)e^{\frac{i\pi(\al+\la-1)}{2}\sigma_{3}}
\left(
\begin{matrix}
1\\
1\\
\end{matrix}
\right)\nonumber\\
&=s^{-\frac{1}{4}\sigma_{3}}\left(I+\mathcal{O}(s^{-\frac{1}{2}})\right)E_{0L}(\hat{z})M_{1}\left(
\begin{matrix}
-iJ_{\al}\left(|\xi|^{\frac{1}{2}}\right)\\
\pi|\xi|^{\frac{1}{2}}J'_{\al}\left(|\xi|^{\frac{1}{2}}\right)
\end{matrix}
\right),
\end{align}
where $M_{1},$ $E_{0L}(\hat{z})$  are given by (\ref{Ro2}) and (\ref{RE111}), respectively.
\par
Set $\xi=n^{2}\phi^{2}(x),$ $x=\frac{u}{16n^{2}}$ and $\xi=n^{2}\phi^{2}(y),$ $y=\frac{{\rm v}}{16n^{2}}$ in (\ref{Ro1}), and inserting these equations into (\ref{RR01}), hence
\begin{align}\label{Ro5}
\frac{1}{4n}K_{n}(\frac{u}{4n},\frac{{\rm v}}{4n}; s)&=\frac{\varphi_{1}(-{\rm v},s)\varphi_{2}(-u,s)-\varphi_{1}(-u,s)\varphi_{2}(-{\rm v},s)}{i2\pi(u-{\rm v})}+\mathcal{O}\left(\frac{1}{n^{2}}\right)\nonumber\\
&=
\frac{J_{\al}(u){\rm v}J'_{\al}({\rm v})-J_{\al}({\rm v})uJ'_{\al}(u)}{2(u-{\rm v})}+\mathcal{O}(s^{-\frac{1}{2}})+\mathcal{O}\left(\frac{1}{n^{2}}\right),\;\;s\rw \infty,\;n\rw \infty,
\end{align}
where the error term is uniform for $u, {\rm v}\in (0,\infty)$ and $t\in(0,c],$ $c$ is a positive and fixed constant. Then (\ref{T9}) derives from (\ref{Ro5}).
\end{proof}

\section{Airy kernel at the soft edge}
{\bfseries The construction of $P^{(1)}(z)$ in the neighbourhood of $z=1.$}
\par
For convenience, we define
\bbq\label{RE115}
\hat{\mu}(z):=\frac{2}{i\pi}\sqrt{\frac{z-1}{z}},\;\; z\in \mathbb{C}\setminus [0,1],
\eeq
and $\hat{\mu}(z)$ such that $\hat{\mu}_{+}(x)=-\hat{\mu}_{-}(x)=\mu(x),$ $x\in (0,1),$ where $\mu(x)$ is given by (\ref{R24}).
\par
We define an auxiliary function $\hat{\phi}(z)$ as follows
\bq\label{R96}
\hat{\phi}(z)=-i\pi\int_{1}^{z}\hat{\mu}(y)dy=-2\int_{1}^{z}\sqrt{\frac{y-1}{y}}dy, \;\; z\in \mathbb{C} \setminus (-\infty,1].
\eq
Note that $\hat{\phi}_{+}(z)$ and $\hat{\phi}_{-}(z)$ purely imaginary on $(0,1)$ and
\bq\label{RE118}
g_{+}(x)-g_{-}(x)=2\hat{\phi}_{+}(x)=-2\hat{\phi}_{-}(x),\;\; x \in (0,1),
\eq
\bbq\label{RE119}
\hat{\phi}_{+}(x)-\hat{\phi}_{-}(x)=i2\pi,\;\; x\in (-\infty,0),
\eeq
\bq\label{RE120}
2\hat{\phi}(x)=2g(x)-4x-\ell_{n},\;\; x\in[1,+\infty).
\eq
where $g(z)$ is given by (\ref{T16}), $\ell_{n}=-2-4\log{2}.$ The above facts can also be found in \cite{QW2008}.
\par
By $(\ref{RE118}),$ $(\ref{RE120}),$ then $(\ref{RE5})$ rewrite as
\bq\label{RE121}
T_{+}(x)=T_{-}(x)\left\{
\begin{array}{ll}
\left(
\begin{matrix}
e^{-2n\hat{\phi}_{+}(x)}&x^{\alpha}\left(x+\frac{t}{4n}\right)^{\lambda}\\
0&e^{-2n\hat{\phi}_{-}(x)}\\
\end{matrix}
\right), & x\in (0,1),
\\
\left(
\begin{matrix}
1&x^{\alpha}\left(x+\frac{t}{4n}\right)^{\lambda}e^{2n\hat{\phi}(x)}\\
0&1\\
\end{matrix}
\right),& x\in \left(1,+\infty\right).
\end{array}
\right.
\eq
$S(z)$ is defined as (\ref{RE13}) and replaced $\phi_{+}(x),$ $\phi_{-}(x)$ by $\hat{\phi}_{+}(x),$ $\hat{\phi}_{-}(x),$ respectively. Then $S(z)$ satisfies the following RH problem.
\par
($S_{a}$) $S(z)$ is analytic in $\C\setminus \{\bigcup_{k=1}^{3}\Sigma_{k}\}\bigcup\left(1, \infty\right),$ see Figure 2.
\par
($S_{b}$) $S_{+}(z)=S_{-}(z)J_{S}$ for $z \in \{\bigcup_{k=1}^{3}\Sigma_{k}\}\bigcup\left(1, \infty\right)$ and the jump $J_{S}$ is given by,
\bq\label{T66}
J_{S}(z)=\left\{
\begin{array}{lll}
\left(
\begin{matrix}
1&0\\
z^{-\alpha}\left(z+\frac{t}{4n}\right)^{-\lambda}e^{-2n\hat{\phi}(z)}&1\\
\end{matrix}
\right),&{\rm for}\; z \in \Sigma_{1}\bigcup\Sigma_{3},\\
\\
\left(
\begin{matrix}
0&z^{\alpha}\left(z+\frac{t}{4n}\right)^{\lambda}\\
-z^{-\alpha}\left(z+\frac{t}{4n}\right)^{-\lambda}&0\\
\end{matrix}
\right),& {\rm for}\;z=x\in (0,1),\\
\\
\left(
\begin{matrix}
1&z^{\alpha}\left(z+\frac{t}{4n}\right)^{\lambda}e^{2n\hat{\phi}(z)}\\
0&1\\
\end{matrix}
\right),& {\rm for}\; z=x\in (1,+\infty),
\end{array}
\right.
\eq
where $\arg \xi\in (-\pi, \pi).$
\par
($S_{c}$) The asymptotic behavior at infinity is
\bbq
S(z)=I+\mathcal{O}\left(z^{-1}\right).
\eeq
\par
($S_{d}$) The asymptotic behavior at $z=1$ is
\bq\label{RE123}
S(z)=\left(
\begin{matrix}
O(1)&O(1)\\
O(1)&O(1)\\
\end{matrix}
\right).
\eq
\par
We seek the local parametrix $P^{(1)}(z)$ in the neighborhood $U(1,r)=\{z\in \mathbb{C}: \left|z-1\right|<r\}$ for small $r>0.$ Moreover, $P^{(1)}(z)$ satisfies the following RH problem.
\par
$(a)$ $P^{(1)}(z)$ is analytic in $U(1,r)\setminus \{\bigcup_{k=1}^{3}\Sigma_{k}\bigcup(1,+\infty)\},$ see Figure 2.
\par
$(b)$ $P^{(1)}(z)$ has the same jump condition as $S(z),$ on $U(1,r)\bigcap \{\bigcup_{k=1}^{3}\Sigma_{k}\bigcup(1,+\infty\},$ see (\ref{T66}).
\par
$(c)$ For $z\in \partial U(1,r),$ $P^{(1)}(z)$ satisfies the following matching condition,
\par
\bq\label{RE124}
P^{(1)}(z)P^{(\infty)-1}(z)=I+\mathcal{O}\left(n^{-1}\right), \;\;n \rw \infty.
\eq
\par
$(d)$ The asymptotic behavior of $P^{(1)}(z)$ at $z=1$ is the same as $S(z)$ in (\ref{RE123}).
\par
Taking the following transformation to constant jump matrices.
\bq\label{RE125}
\hat{P}^{(1)}(z)=E_{1}^{-1}(z)P^{(1)}(z)e^{n\hat{\phi}(z)\sigma_{3}}\left(z+t/4n\right)^{\frac{\la}{2}\sigma_{3}}z^{\frac{\al}{2}\sigma_{3}},
\eq
where $E_{1}(z)$ is an invertible analytic matrix function in $U(1,r).$ $\hat{P}^{(1)}(z)$ fulfills the following RH problem.
\par
$(a)$ $\hat{P}^{(1)}(z)$ is analytic in $U(1,r)\setminus \{\bigcup_{k=1}^{3}\Sigma_{k}\bigcup(1,+\infty)\}.$
\par
$(b)$ $\hat{P}^{(1)}(z)$ satisfies the following jump condition,
\bq\label{RE126}
\hat{P}_{+}^{(1)}(z)=\hat{P}_{-}^{(1)}(z)\left\{
\begin{array}{lll}
\left(
\begin{matrix}
1&0\\
1&1\\
\end{matrix}
\right),&{\rm for}\; z \in U(1,r)\cap\{\Sigma_{1}\cup\Sigma_{3}\},\\
\\
\left(
\begin{matrix}
0&1\\
-1&0\\
\end{matrix}
\right),& {\rm for}\;z=x\in U(1,r)\cap(0,1),\\
\\
\left(
\begin{matrix}
1&1\\
0&1\\
\end{matrix}
\right),& {\rm for}\; z=x\in U(1,r)\cap(1,+\infty).
\end{array}
\right.
\eq
\par
It is well-known that the Airy function has been successful applied to construct the local parametrix in \cite{DeiftBook, DKV1999}, see also \cite{V2007}. Some more information, see \cite{OLC2010}. With the jump condition of $\hat{P}^{(1)}(z)$ in $(\ref{RE126}),$ it is natural to consider the following RH problem for $\widetilde{P}^{(1)}(z)$ which can be constructed by the Airy function and its derivatives.
\par
$(a)$ $\widetilde{P}^{(1)}(z)$ is analytic in $z\in \mathbb{C}\setminus \cup_{k=1}^{4}\Sigma''_{k},$ see Figure 9.
\par
$(b)$ $\widetilde{P}^{(1)}(z)$ satisfies the jump condition
\bbq\label{RE127}
\widetilde{P}^{(1)}_{+}(z)=\widetilde{P}^{(1)}_{-}(z)\left\{
\begin{array}{lll}
\left(
\begin{matrix}
1&0\\
1&1\\
\end{matrix}
\right), & z\in \Sigma''_{1}\cup\Sigma''_{3},\\
\\
\left(
\begin{matrix}
0&1\\
-1&0\\
\end{matrix}
\right),& z\in \Sigma''_{2},\\
\\
\left(
\begin{matrix}
1&1\\
0&1\\
\end{matrix}
\right),&z\in \Sigma''_{4},
\end{array}
\right.
\eeq
\par
$(c)$ For $z\rw \infty,$
\bq\label{RE128}
\widetilde{P}^{(1)}(z)=z^{-\frac{1}{4}\sigma_{3}}\frac{1}{\sqrt{2}}\left(
\begin{matrix}
1&1\\
-1&1\\
\end{matrix}
\right)
\left(I+\mathcal{O}\left(z^{-\frac{3}{2}}\right)\right)e^{-\frac{i\pi}{4}\sigma_{3}}e^{-\frac{2}{3}z^{\frac{3}{2}}\sigma_{3}},
\eq
where $\arg\xi\in(-\pi, \pi).$
\begin{center}

\begin{tikzpicture}
\begin{scope}[line width=2pt]
\draw[->,>=stealth] (0,0)--(1.5,0);
\draw[-] (0,0)--(4.6,0);

\draw[->,>=stealth] (-5,0)--(-2,0);
\draw[-] (-3,0)--(0,0);
\draw[-] (1,0)--(2,0);
\draw[->,>=stealth] (-1.733,2.1)--(-1.3,1.575);
\draw[-] (-2.6,3.15)--(0,0);
\draw[->,>=stealth] (-1.733,-2.1)--(-1.3,-1.575);
\draw[-] (-2.6,-3.15)--(0,0);
\node[below] at (0,0) {$O$};
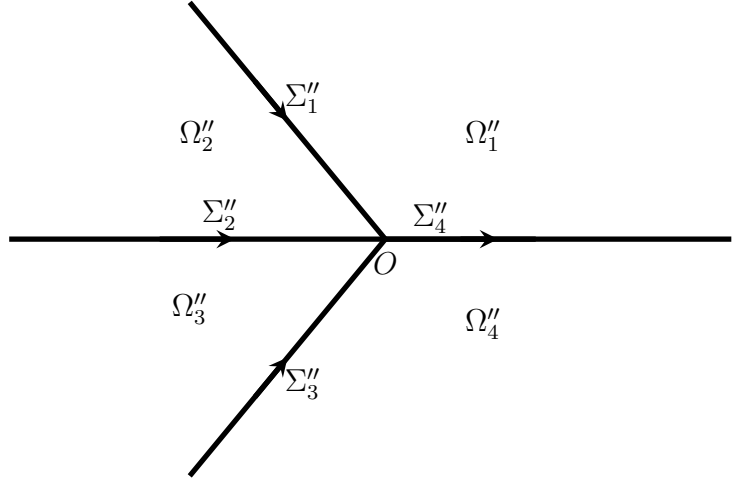
\node[below] at (-.8,-3.5) {Fig.9. Contours $\mathbb{C}\setminus \displaystyle\cup_{j=1}^{4}\Sigma''_{j},$ and regions $\Omega''_{j}, j=1,\ldots,4.$ };

\node[above] at (-1.1,-2.3) {{$\Sigma''_{3}$}};
\node[above] at (-2.2,-.05) {{$\Sigma''_{2}$}};
\node[above] at (-1.1,1.5) {{$\Sigma''_{1}$}};
\node[above] at (.6,-.1) {{$\Sigma''_{4}$}};

\node[above] at (1.3,1) {{$\Omega''_{1}$}};
\node[above] at (-2.5,1) {{$\Omega''_{2}$}};
\node[above] at (-2.6,-1.3) {{$\Omega''_{3}$}};
\node[above] at (1.3,-1.5) {{$\Omega''_{4}$}};
\end{scope}
\end{tikzpicture}
\end{center}
\par
A solution to the above RH problem for $\widetilde{P}^{(1)}(z)$ can be constructed by the Airy function and its derivatives as follows
\bq\label{RE1130}
\widetilde{P}^{(1)}(z)=M_{2}\left\{
\begin{array}{llll}
\left(
\begin{matrix}
Ai(z)&Ai(\omega^{2}z)\\
Ai'(z)&\omega^{2}Ai'(\omega^{2}z)\\
\end{matrix}
\right)e^{-i\frac{\pi}{6}\sigma_{3}}, &z \in \Omega''_{1},\\
\\
\left(
\begin{matrix}
Ai(z)&Ai(\omega^{2}z)\\
Ai'(z)&\omega^{2}Ai'(\omega^{2}z)\\
\end{matrix}
\right)e^{-i\frac{\pi}{6}\sigma_{3}}
\left(
\begin{matrix}
1&0\\
-1&1\\
\end{matrix}
\right)
, &z \in \Omega''_{2},\\
\\
\left(
\begin{matrix}
Ai(z)&-\omega^{2}Ai(\omega{z})\\
Ai'(z)&-Ai'(\omega{z})\\
\end{matrix}
\right)e^{-i\frac{\pi}{6}\sigma_{3}}
\left(
\begin{matrix}
1&0\\
1&1\\
\end{matrix}
\right)
, &z \in \Omega''_{3},\\
\\
\left(
\begin{matrix}
Ai(z)&-\omega^{2}Ai(\omega{z})\\
Ai'(z)&-Ai'(\omega{z})\\
\end{matrix}
\right)e^{-i\frac{\pi}{6}\sigma_{3}}
, &z \in \Omega''_{4},\\
\end{array}
\right.
\eq
where $M_{2}=\sqrt{2\pi}e^{-i\frac{\pi}{12}},$ the regions $\Omega''_{j}, j=1,\ldots,4$ illustrated in Figure 9.
\par
We construct $\hat{P}^{(1)}(z)$ by defining $\hat{P}^{(1)}(z):=\widetilde{P}^{(1)}(g_{n}(z)),$ where $g_{n}(z): U(1,r)\rw g_{n}(U(1,r))$ with $g_{n}(1)=0,$ is an appropriate biholomorphic map. To matching the exponential term in $(\ref{RE125})$ with the asymptotic expansion of $\widetilde{P}^{(1)}(z)$ in $(\ref{RE128}),$ let
\bq\label{RE129}
-\frac{2}{3}g_{n}^{\frac{2}{3}}(z)=n\hat{\phi}(z),\;\;z\in U(1,r),
\eq
by $(\ref{R96}),$ it rewrites as
\bq\label{RE130}
g_{n}(z)=\left(3n\int_{1}^{z}\sqrt{\frac{x-1}{x}}dx\right)^{\frac{2}{3}},
\eq
and its Maclaurin expansion at $z=1,$
\bq\label{R63}
g_{n}(z)=(2n)^{\frac{2}{3}}(z-1)\left(1-\frac{1}{5}(z-1)+\frac{17}{157}(z-1)^{2}+\mathcal{O}((z-1)^3)\right).
\eq
\par
By (\ref{RE124}), (\ref{RE125}), (\ref{RE128}) and (\ref{RE129}), then the explicit $E_{1}(z)$ and $P^{(1)}(z)$ are given by
\bq\label{RE132}
P^{(1)}(z)=E_{1}(z)\widetilde{P}^{(1)}(g_{n}(z))e^{-n\hat{\phi}(z)}\left(z+t/4n\right)^{-\frac{\la}{2}\sigma_{3}}z^{-\frac{\al}{2}\sigma_{3}},
\eq
and
\bq\label{RE133}
E_{1}(z)=P^{(\infty)}(z)z^{\frac{\al}{2}\sigma_{3}}\left(z+t/4n\right)^{\frac{\la}{2}\sigma_{3}}e^{i\frac{\pi}{4}\sigma_{3}}\frac{1}{\sqrt{2}}\left(
\begin{matrix}
1&-1\\
1&1\\
\end{matrix}
\right)g_{n}^{\frac{1}{4}\sigma_{3}}(z),
\eq
where $P^{(\infty)}(z),$ $\widetilde{P}^{(1)}(z)$ are given by (\ref{R14}) and (\ref{RE1130}), respectively.
\par
By (\ref{R14}), (\ref{RE132}) and (\ref{RE133}), the matching condition (\ref{RE124}) is easy to verify.
\par
For small $r,$ we define
\bq\label{RE134}
R_{1}(z,t)=\left\{
\begin{array}{ll}
S(z,t)P^{(\infty)-1}(z), & \left|z-1\right|>r,\\
\\
S(z,t)P^{(1)-1}(z,t),& \left|z-1\right|<r.\\
\end{array}
\right.
\eq
With $P^{(\infty)}(z),$ $P^{(1)}(z)$ are in (\ref{R14}), (\ref{RE132}), respectively, after some calculations, one has,
\bq\label{RE135}
J_{R_{1}}(z)=\left\{
\begin{array}{lll}
I+\mathcal{O}\left(n^{-1}\right),& z \in \Sigma''_{2}\cap U(1,r),\\
\\
I+\mathcal{O}\left(n^{-1}\right),& z\in \partial U(1,r),\\
\\
I+\mathcal{O}\left(e^{-cn}\right),& z\in U(1,r)\cap \Sigma''_{1}\cap\Sigma''_{3}\cap\Sigma''_{4},
\end{array}
\right.
\eq
where $c$ is a positive constant. By a similar argument in section 3.7, we have,
\bq\label{RE136}
R_{1}(z)=I+\mathcal{O}\left(n^{-1}\right),
\eq
where the error term is uniform for bounded $z$ and $t$ in compact subsets of $(0,\infty).$
\\
\par
{\bfseries Proof of Theorem 4}
\par
\begin{proof}
From (\ref{RW2}), (\ref{RE13}) and (\ref{RE134}), one takes the inverse transformations from $Y$ to $R_{1},$ and combining with (\ref{RE132}), (\ref{RE133}) and (\ref{RE120}), then the explicit expression of $Y_{+}(x)$ in the interval $(1,1+r)$ is given by
\begin{align}\label{RE137}
Y_{+}(4nx)\left(
\begin{matrix}
1\\
0\\
\end{matrix}
\right)&=(4n)^{\left(n+\frac{\al+\la}{2}\right)\sigma_{3}}e^{-\frac{n\ell}{2}\sigma_{3}}R_{1}(x)E_{1}(x)\widetilde{P}^{(1)}\left(g_{n}(x)\right)
(w(4nx, t))^{-\frac{1}{2}\sigma_{3}}
\left(
\begin{matrix}
1\\
0\\
\end{matrix}
\right)\nonumber\\
&=(4n)^{\left(n+\frac{\al+\la}{2}\right)\sigma_{3}}e^{-\frac{n\ell}{2}\sigma_{3}}R_{1}(x)E_{1}(x)\widetilde{P}^{(1)}\left(g_{n}(x)\right)\left(
\begin{matrix}
1\\
0\\
\end{matrix}
\right)(w(4nx, t))^{-\frac{1}{2}},
\end{align}
which is also can rewrite as
\bq\label{RE139}
Y_{+}(4nx)(w(4nx, t))^{\frac{1}{2}}\left(
\begin{matrix}
1\\
0\\
\end{matrix}
\right)
=(4n)^{\left(n+\frac{\al+\la}{2}\right)\sigma_{3}}e^{-\frac{n\ell}{2}\sigma_{3}}R_{1}(x)E_{1}(x)\widetilde{P}^{(1)}\left(g_{n}(x)\right)\left(
\begin{matrix}
1\\
0\\
\end{matrix}
\right),
\eq
where $\widetilde{P}^{(1)}(z),$ $g_{n}(z)$ are given by $(\ref{RE1130}),$ $(\ref{RE130}),$ respectively.
\par
With the aid of (\ref{RE139}), the equation (\ref{RE138}) can be rewritten as
\bq\label{RE140}
4nK_{n}(4nx,4ny; t)=\frac{\left(
\begin{matrix}
0&1\\
\end{matrix}
\right)\widetilde{P}^{(1)-1}\left(g_{n}(y)\right)E^{-1}(y)R_{1}^{-1}(y)R_{1}(x)E(x)\widetilde{P}^{(1)}\left(g_{n}(x)\right)\left(
\begin{matrix}
1\\
0\\
\end{matrix}
\right)}{i2\pi(x-y)}.
\eq
\par
From (\ref{R63}) and $z\in \Omega''_{2},$ see Figure 9, variables $x$ and $y$ re-scaled as $x=1+(2n)^{-\frac{2}{3}}u,$ $y=1+(2n)^{-\frac{2}{3}}{\rm v},$ where $u, {\rm{v}}\in (-\infty,0).$
$E_{1}(z)$ is analytic and bounded in $U(1,r)$ for small $r$ and its explicit expression given by (\ref{RE133}), after some calculations, one finds,
\bq\label{RE143}
E_{1}^{-1}(y)E_{1}^{-1}(x)=I+\mathcal{O}\left(x-y\right)=I+(u-{\rm v})\mathcal{O}\left(n^{-\frac{2}{3}}\right).
\eq

From (\ref{RE136}), it follows that
\bq\label{RE146}
R_{1}^{-1}(y)R_{1}(x)=I+\left(x-y\right)\mathcal{O}\left(n^{-1}\right)=I+\left(u-{\rm v}\right)\mathcal{O}\left(n^{-\frac{5}{3}}\right).
\eq
\par
With (\ref{R63}) and $x=1+(2n)^{-\frac{2}{3}}u,$ one finds,
$
g_{n}(x)=u\left(1-\frac{1}{5}\left(2n\right)^{-\frac{2}{3}}u+\mathcal{O}\left(n^{-\frac{4}{3}}\right)\right),
$
and
\bq\label{RE142}
Ai\left(g_{n}(x)\right)=Ai(u)+\mathcal{O}\left(n^{-\frac{2}{3}}\right),
\eq
similar estimations are also valid for $g_{n}(y)$ and $Ai\left(g_{n}(y)\right).$
\par
The above asymptotic expansions (\ref{RE143}), (\ref{RE146}) and (\ref{RE142}) are uniform for $u$ and ${\rm v}$ in compact subsets of $(-\infty, 0).$ Inserting (\ref{RE1130}), (\ref{RE143}), (\ref{RE146}), and (\ref{RE142}) into (\ref{RE140}), one finds,
\bq\label{T22}
2(2n)^{\frac{1}{3}}K_{n}(4n+2(2n)^{\frac{1}{3}}u,4n+2(2n)^{\frac{1}{3}}{\rm v}; t)=\frac{Ai(u)Ai'({\rm v})-Ai({\rm v})Ai'(u)}{u-{\rm v}}+\mathcal{O}\left(n^{-\frac{2}{3}}\right),
\eq
and this asymptotic expansion is uniform for $u,$ ${\rm v}$ and $t$ in compact subsets of $(-\infty, 0).$ Then (\ref{T10}) derives from (\ref{T22}).
\end{proof}

{\bf Acknowledgement.}
\par
This work of En-Gui Fan was  supported by the National Science Foundation of China under Project No.11671095.
\\

     \end{document}